\documentclass[%
 reprint,
 amsmath,amssymb,
 aps,
nofootinbib]{revtex4-1}

\usepackage{graphicx}
\usepackage{dcolumn}
\usepackage{bm}
\usepackage[symbol]{footmisc}

\usepackage{amsmath}
\usepackage{xcolor}
\usepackage{float}
\usepackage{multirow}
\usepackage{chemformula}

\begin{document}

\preprint{APS/123-QED}

\title{Changes in Dislocation Punching Behavior Due to Hydrogen-Seeded Helium Bubble Growth in Tungsten}

\author{Peter Hatton}
\affiliation{Material Science and Technology Division, Los Alamos National Lab, Los Alamos, 87545, NM, USA. }

\author{Danny Perez}
\affiliation{Theoretical Division, Los Alamos National Lab, Los Alamos, 87545, NM, USA}

\author{Blas Pedro Uberuaga}
\affiliation{Material Science and Technology Division, Los Alamos National Lab, Los Alamos, 87545, NM, USA}

\date{\today}

\begin{abstract}
The accumulation of gas atoms in tungsten is a topic of long-standing interest to the plasma-facing materials community due the metal's use as a divertor material in some tokamak fusion reactors. The nucleation and growth of He/H gas bubbles (along with their isotopes) can result from impinging fluxes of these gases which give rise to damage at the W divertor surface. The inclusion of He \emph{or} H in W has been studied extensively by the community, finding that He bubbles modify the surface through periodic dislocation punching and bursting mechanisms while H bubbles impact the metal through plastic-strain induced material failure. However, the mechanisms which are present during the combined flux of both He and H is not well-studied atomistically. Motivated by this, an atomistic modeling study is conducted using molecular dynamics to assess the behavior of mixed concentration He:H bubbles in W. We find that the introduction of H into a growing He bubble results in a dramatic change in the nature and presence of dislocation loops which are typically generated via dislocation punching in over-pressurized He bubbles. Most notably, at high H concentrations, there is a switchover in energetic favorability from glissile 1/2$<$111$>$ dislocations to sessile $<$100$>$ dislocations. This thermodynamic crossover could imply a significant reduction in W surface morphology changes than with pure He bubbles and, additionally, we show this to have implications on the trapping of H in the bubbles and their associated dislocations. 
\end{abstract}

\maketitle

\section{Introduction}

The accumulation of gas atoms in tungsten has been the topic of significant research as it is a critical issue for the durability of plasma-facing materials in fusion reactors. During fusion reactor operation, the expected high flux of gas atoms from the plasma to the W divertor will inevitably lead to their inclusion in the metal. These incident atoms primarily consist of Helium (He) and Hydrogen (H), including the H isotopes Deuterium (D) and Tritium (T), which act as fuel in the fusion reaction. Their inclusion in W can detrimentally alter its thermal and mechanical properties \cite{Chen2020,Iwakiri2000}. Further, the implantation of gases under fusion conditions significantly modifies the surface morphology of the material, increasing the potential for release of high Z dust, impacting the energy efficiency of the plasma. Finally, there are concerns about the retention of radioisotopes such as T that impact the environmental handling of materials. The mitigation of these mechanisms is of paramount importance to ensuring the longevity and safety of reactor operation. 

The inclusion of He \emph{or} H (and its isotopes) in W has been studied extensively by the community. Computational work indicates that He inclusion leads to the formation of He bubbles which, during their growth in bulk W, generate glissile $1/2<$111$>$ dislocation loops which, at a certain size, emit from the bubble to a free surface \cite{Sefta2013,Iwakiri2000,Fikar2017,Sandoval2018}. This process acts as a pressure release mechanism for the bubble and results in the creation of surface islands which act to progressively modify the W surface \cite{Hammond2020,Hammond2019,Fikar2017, Fikar2015}. Further, the bubble motion toward the surface as a result of progressive dislocation nucleation and punching or due to Frenkel pair induced transport \cite{Perez2017} can cause surface rupture leading to out-gassing of the He to the vacuum and further surface modification \cite{Sandoval2015,Ito2014}. Through multi-scale modeling efforts, these mechanisms have been strongly correlated to the creation of the well-known W surface \lq fuzz,' \cite{Dasgupta2019,Ueda2013,Ueda2014,Wang2017_a,lasa2013,Lasa2014,Chen2022} degrading the surface and producing nano-tendrils of W which emanate from its surface that could break off, impacting plasma performance. Somewhat counter to this are He bubbles growing at W GBs, as can be the case in polycrystalline W, due to the strong segregation of defects to a GB region \cite{ElAtwani2017,ElAtwani2015_a,Bai2010,Uberuaga2015,Bai2013,Xiao2016}. In this scenario the He bubble can instead act to transform the density of the GB phase as a pressure releasing mechanism rather than from forming surface altering dislocation loops, but this conclusion may depend 
on the character of the specific GB \cite{Hatton2024,Frolov2018}.

Studies of H (D,T) inclusion in W commonly find the generation of surface blisters caused by the agglomeration of H gas into large subsurface bubbles due to their fast kinetics \cite{Lu2014,Liu2009,Johnson2010,Heinola2010,Frauenfelder1969} and thermodynamic driving force to accumulate at W defects \cite{Johnson2010,Duan2010,Liu2009_a,Heinola2010,Ohsawa2010,Ohsawa2012,Lu2014}. This induces plastic strain at the surface \cite{Chen2020,Ye2003,Tanabe2014,Lu2014} resulting in blisters, an effect seen in other materials subjected to a plasma which can, in some cases, result in surface rupture and dramatic out-gassing \cite{Hatton2019,Hatton2020}. It has also been suggested that GBs can have a trapping effect on H \cite{Yu2014,Wang2017} which results in complex effects on the GB mobility and recrystallization of W \cite{Mathew2022} thereby impacting its mechanical properties \cite{Shi2023,Ding2022}.

Experimental studies of a combined He \emph{and} H flux indicate increased H retention in areas of the sample where He is present and an overall increase in H retention compared with a pure H exposure, implying some trapping enhancement caused by interactions between H and He \cite{Nishijima2005}. This effect has been confirmed through computational work indicating that H preferentially decorates the periphery of the He bubble and is strongly bound to this position \cite{Bergstrom2017}. Further, experimental studies which analyse surface damage as a result of combined gas fluxes provide some evidence that, compared to cases with only He \emph{or} H fluxes, surface damage is significantly reduced \cite{Gao2014,Miyamoto2009,Alimov2009,Miyamoto2011}. That is, in the combined case there is no evidence of surface blistering (as was the case with pure H implantation) nor surface fuzz (as with pure He implantation), indicating some proposed \lq cancelling' effect when the two gasses are combined \cite{Gao2014,Miyamoto2009,Alimov2009,Miyamoto2011}. 
 
Previous computational studies of mixed He/H bubbles have concentrated on static properties, but did not shed light on the development of dislocation loops around the gas bubbles, an effect previously shown to be important when assessing surface W damage as a result of gas influx \cite{Sefta2013,Iwakiri2000,Fikar2017,Sandoval2018}. The goal of the present study is precisely to investigate the effect that growing He:H bubbles have on the surrounding W matrix and the resulting character of the dislocation networks. This will allow the elucidation of the mechanisms which impact W surface damage in environments of combined He:H gas flux. More specifically, in section \ref{sec:1} we use molecular dynamics (MD) to grow mixed He:H bubbles with a range of gas ratios. We analyse the behavior of the H in these bubbles and characterize the dislocation network which arises from these mixed gas bubbles. We then complete static calculations of dislocation loops which may result from He bubbles to understand how their formation energy is impacted by the presence of H. In section \ref{sec:3}, we analyse a specific case of an equal ratio He:H bubble close to the W surface to understand its behavior and impact on H retention and surface damage. Finally, in section \ref{sec:Disc} we discuss the implications of the changes in SIA and dislocation behavior induced by the additional presence of H on W surface microstructural evolution.

\section{Methodology}

This work uses the Molecular Dynamics (MD) code LAMMPS \cite{LAMMPS2022} with atomic interactions modeled using the W-H interatomic potential proposed by Gao \emph{et al.} \cite{Wang2017_b} which itself uses Marinica's EAM potential \cite{Marinica2013} for W-W interactions. W-He and He-He interactions are modelled using the Juslin-Wirth potential set \cite{Juslin2013}.

MD simulations of He:H bubble growth are conducted within a W supercell of size $127 \times 127 \times 60$ Å$^3$ with $20$ Å of vacuum above and below the supercell to create two $<$100$>$ free surfaces in the z-direction. The supercell initially contains 64,000 atoms with periodic boundaries in each direction. A W atom is removed from the center of the supercell and 4 gas atoms are initially placed within this W vacancy site to seed the bubble with the desired He:H ratio, the structure is minimized, and the system is thermalized by running MD for 10 ps at 1000 K. MD is then conducted at 1000 K and a subsequent He or H is inserted randomly into a sphere of radius 1.4 Å centered on the initial W vacancy at a rate of 10$^{11}$ gas/s (or, every 10 ps). Due to H moving rapidly at these temperatures inside the bubble, it was necessary to use an MD timestep of 0.1 fs to avoid instability in the dynamics.

To understand the impact of H on the thermodynamic stability of different dislocation loop structures, static calculations of dislocation loop's formation energies with and without the presence of H along their dislocation cores are also completed. Loops are created in a 150 Å cubic box with periodic boundary conditions in each direction initially containing 250,000 atoms. Loops with Burgers vectors along 1/2$<$111$>$ and $<$100$>$ are created of various sizes using a script which takes advantage of a multitude of modifiers within Ovito \cite{Stukowski2010}. The formation energy ($E_f$) is calculated using the standard formula,

\begin{equation}\label{eqn:CleanForm}
E_f = E_{\text{loop}} - N_W E_c^W,
\end{equation}

\noindent where $E_{\text{loop}}$ is the energy of the system containing the loop and $N_W$ is the number of W atoms in this system. $E_c^W$ is then the cohesive energy of W, calculated with this potential to be 8.87 eV/atom. The formation energy of these dislocation loops when the core is saturated with H ($E^H_f$) is then calculated as

\begin{equation}\label{eqn:HForm}
E^H_f = E^H_{\text{loop}} - N_W E_c^W - N_H E_f^{H_i -> W}.
\end{equation}

\noindent Here $E^H_{\text{loop}}$ is the energy of the system containing the dislocation loop with $N_H$ H atoms along the core and $E_f^{H_i -> W}$ is the formation of a single H interstitial in clean, bulk W. A low energy position of H on each dislocation core is found by trialling 500 initial H positions within a 2 Å diameter toroid around the dislocation core and an energy minimization is conducted for each of these candidate positions. The formation energy is then calculated in this way for many different H counts at the core. After some number of H atoms are added to the core through this process, it will reach a saturation point; this saturation point is declared when adding more H to the core causes an increase in $E^H_f$ for all of the trialled H positions. Further, initial H positions are rejected if, as a result of their placement and minimization, the dislocation moves more than 2 Å or if the introduction of H atoms results in the dislocation no longer being detected; this can happen if the initial position of a H atom is too close to a W atom and induces an unphysical reconstruction leading to the collapse of the dislocation loop. The use of $H_i$ as a reference for H means that this energy assumes H atoms are coming in from the bulk as interstitial species and are then binding to the loop. The visualization and collection of dislocation and interstitial/vacancy data during MD and static simulations is completed using DXA and Wigner-Seitz analysis as implemented in Ovito \cite{Stukowski2010}.

\section{Results}

\subsection{Dislocation Loop Modification in the Presence of H}\label{sec:1}

\begin{figure*}
\includegraphics[width=\linewidth]{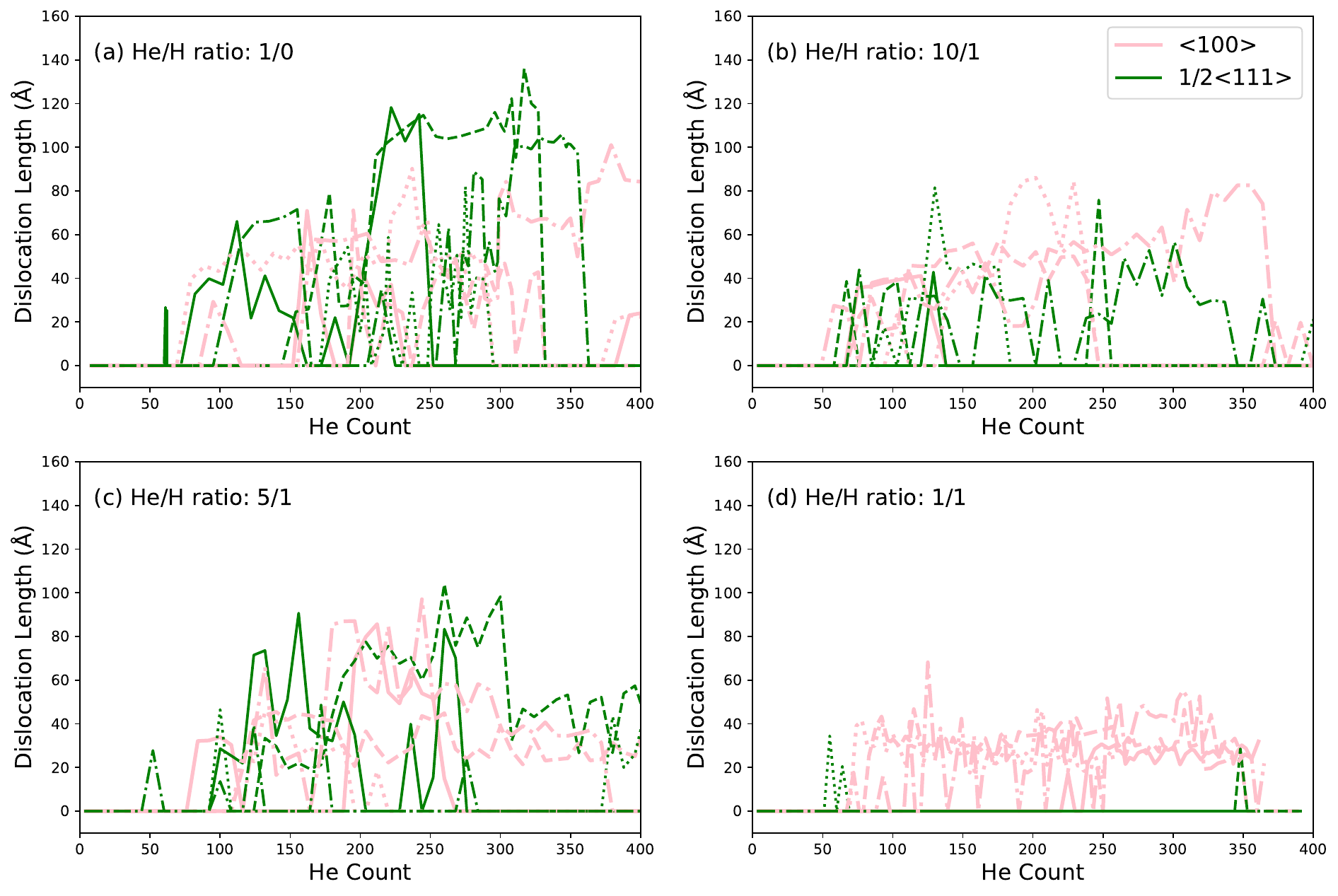}
\caption{\label{fig:DislocationVsCount} Dislocation length versus bubble size for different He:H ratios. Each panel contains 4 different realizations, represented by different types of dashed lines, of the dynamics each having distinct random velocity seeds.
}
\end{figure*}

Mixed bubbles are grown with direct MD at different He:H ratios from 1:0 (pure He) to a 1:1 He:H ratio and the length of different dislocation types with increasing bubble size are extracted from the simulations which can be seen in figure \ref{fig:DislocationVsCount}. Firstly, the case with only He (ratio of 1:0) shows the typical nucleation and growth of dislocation loops with a mixture of 1/2$<$111$>$ (green) and $<$100$>$ (pink) burgers vectors induced by Frenkel pair creation by the growing bubble. The mixture of loop types implies there is some competition between generating each type of dislocation loop at the bubble. This is consistent with experimental findings in W which find a mixture of loop types present in damaged tungsten \cite{Yi2015,Yi2016}. 

The identification of dislocation loop type is key to understanding how the surface morphology will be affected by the growing bubble. Dislocation loops with burgers vector of 1/2$<$111$>$ are known to be glissile in W, meaning they are able glide through the pristine metal \cite{Hull2001}. In contrast, loops with $<$100$>$ burgers vectors are sessile meaning they are static in W and cannot migrate in the crystal \cite{Hull2001}. Therefore, the presence of 1/2$<$111$>$ dislocation loops in the pure He bubble growth cases results in the deposition of W atoms onto the surface of the simulation cell through the glissile motion of the loops once they detach from the bubble surface. This is characteristic of the dislocation loop punching mechanism identified in previous atomistic and continuum models/studies where 1/2$<$111$>$ loops will periodically nucleate, grow in size, and then punch to the free surface \cite{Setyawan2023}.

In figures~(\ref{fig:DislocationVsCount})b-c (ratios of 10:1 and 5:1), we consider the effect of adding a low concentration of H to the bubble. Similar to the pure He case, we also see evidence of a mixture of dislocation types and evidence of the loop punching process which result in an average number of adatoms on the surface at the end of the simulation to be 110 for the 10:1 case and 90 for the 5:1 compared to 161 in the 1:0 case. This indicates a softening of the loop punching mechanism with H present which is consistent with the observed reduction in the total dislocation lengths created during these simulations.

\begin{figure}
\includegraphics[width=\linewidth]{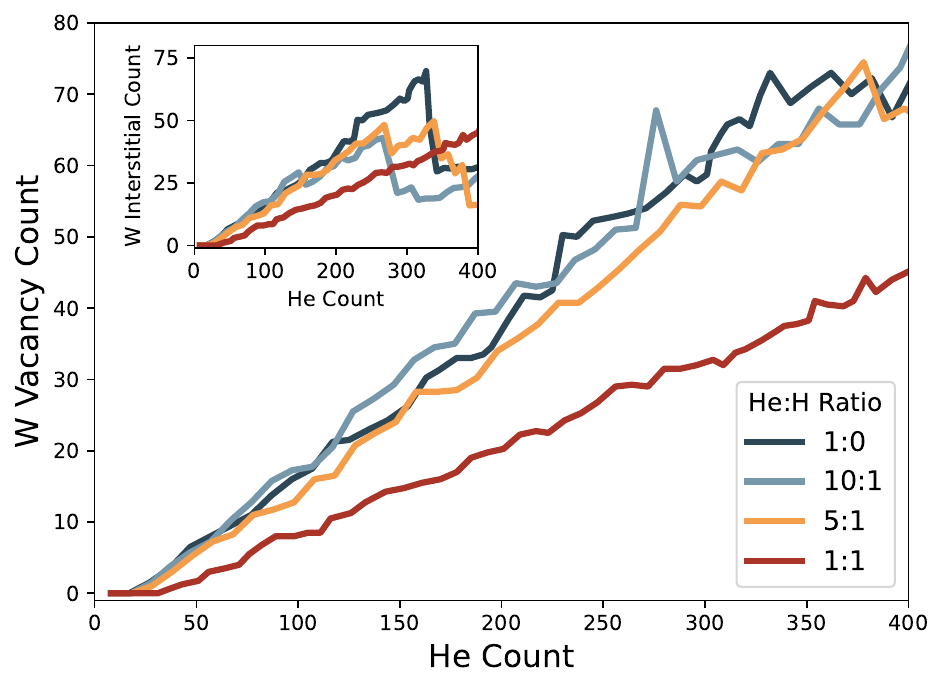}
\caption{\label{fig:WInts} W vacancy count against He count in the bubble averaged over the realizations for each He:H ratio. Inset shows interstitial count in which W atoms which reach the surface are not included in the count.}
\end{figure}

In bubbles grown with high H concentration (ratio of 1:1) we observe a large decrease in the overall length and occurrence of dislocation loops, shown in figure \ref{fig:DislocationVsCount}(d). Additionally, 1/2$<$111$>$ dislocations are almost non-existent in this case. Only in one of the four simulations performed for this ratio do we see significant surface modification where a number of adatoms were able to reach the surface. This was a case where the bubble grew preferentially towards the free surface and will be discussed in detail in section \ref{sec:3}. However, in general, surfaces were left undefected with bubbles grown with equal parts He and H. This is because $<$100$>$ dislocation loops, which still form in these bubble conditions, are sessile meaning they cannot glide to the free surface and instead their interstitials stay attached to the bubble. This implies that there is some He:H ratio in a bubble ($\simeq$ 1:1) which causes a drastic altering in the type of dislocation loops that are favorable, shifting from predominantly glissile to sessile structures, and therefore eliminating the generation of surface adatoms caused by the growing bubble.

\begin{figure}
\includegraphics[width=\linewidth]{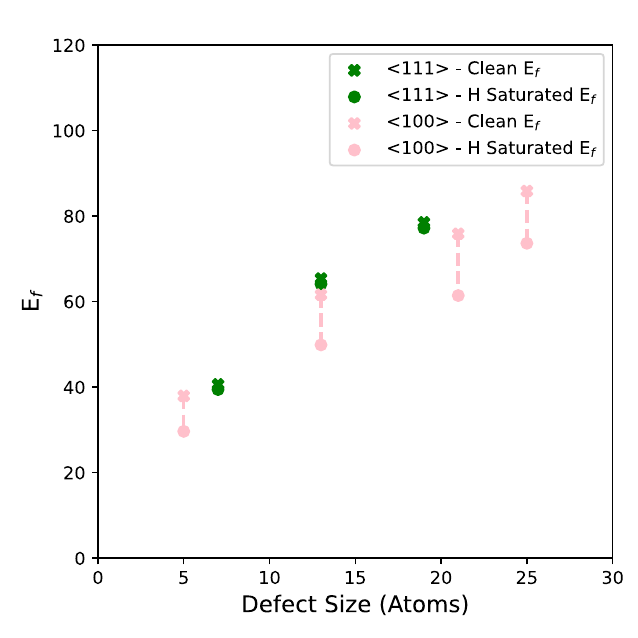}
\caption{\label{fig:FormEn_Pot} Formation energy of isolated dislocation loops of different defect sizes with and without H saturation on the loop.}
\end{figure}

Two effects have therefore been noted from this analysis. Firstly, there is an overall decrease in the total length and occurrence of dislocation loops with increasing H concentration in the growing bubble, we refer to this as a reduction in dislocation activity. Secondly, at high H concentration, the nature of the dislocations that do form also change, that is, there is a significant decrease in the presence of 1/2$<111>$ loops leading almost to their elimination in the 1:1 case.

The overall reduction in dislocation lengths when introducing H can be explained by observing Fig \ref{fig:WInts} which shows the number of W vacancies created during the simulations (averaged over the 4 realizations of each He:H ratio). From this we can infer that at low-mid H concentrations (10:1 and 5:1) the total number of W vacancies (and therefore interstitials) which are created is similar to the pure He case. However, from Fig 1(b,c) we observe less dislocation activity. This must imply that there is an increased number of non-loop W SIAs present and that H is acting - through a W-H binding action - to inhibit the formation of loops. Further, at high H concentration (1:1) there is a significant reduction in the number of W SIA's created at a given He count. In this case the W-H binding is acting to increase the energy barrier and/or the $\Delta E$ for the creation of a Frenkel pairs hence inhibiting the SIA creation mechanisms altogether.

Additionally, Fig \ref{fig:WInts}(inset) shows a count of the current SIAs present within W during the simulations. The large drops in these values correspond to 1/2$<$111$>$ dislocations punching to the surface, depositing their interstitials. From this it is clear that the net effect of the higher concentration of H is the significant softening and almost elimination of the dislocation punching events and therefore the reduced roughening of the W surface compared to purely He bubbles.

However, the overall reduction in SIA's caused by H does not explain the observed change in dislocation loop character from primarily glissile 1/2$<$111$>$ to sessile $<$100$>$ loops. To explain this we calculate the formation energy of isolated dislocation loops with and without H to determine how the loop formation energy depends on the presence of H. It has been established in the literature \cite{DeBacker2017,Bergstrom2017}, and indeed we have seen in these simulations, that H can become bound to dislocation loops. Therefore, the hypothesis is that the presence of H may have some effect on the stability of the dislocation loops.

The formation energy of clean, isolated dislocation loops can be seen in figure \ref{fig:FormEn_Pot} (crosses) and are calculated using equation \ref{eqn:CleanForm}. With this interatomic potential, and without H, the formation energy of 1/2$<$111$>$ loops and $<$100$>$ loops is nearly degenerate, which is consistent with the behavior exhibited in figure \ref{fig:DislocationVsCount}(a) (the 1:0 ratio case) showing a mixture of both dislocation types. This is also consistent with experimental studies which find both loop types present upon Frenkel pair inducing damage in W \cite{Yi2015,Yi2016}.

The dislocations are then saturated with H as described in the methodology and the formation energies re-calculated using equation \ref{eqn:HForm}. The resulting loop formation energies can also be seen in figure \ref{fig:FormEn_Pot} (circles). The amount of H required to saturate each loop strongly depends on the nature of the loop. 1/2$<$111$>$ loops saturate with only $\simeq$ 5 H on the core - relatively independent of size - whereas on $<$100$>$ loops the saturation point increases steadily from $\simeq$ 5-20 as the size of the loop increases. At the loop saturation point there is a significant decrease in the formation energy of $<$100$>$ loops at each loop size. Counter to this is the decorated 1/2 $<$111$>$ loop which exhibit a negligible decrease in energy as H is added to the core. Thus, H transforms the energy landscape such that, while in pure W the energy of these two loops is nearly degenerate, upon the addition of H there is a significantly lower energy associated with forming the $<$100$>$ loop type and the simulation will be heavily biased away from forming 1/2$<$111$>$ loops.

In summary, during the growth of mixed He:H bubbles the preference for the formation of sessile $<$100$>$ loops, alongside the overall decrease in SIA production, leads to the decrease in surface morphology alterations caused by ejected W interstitials in the simulations. This change is a consequence of greater binding of H to $<$100$>$ loops. This could have significant impact on the evolution of the W divertor surface under fusion reactor conditions.

\subsection{Mixed He:H Bubbles in a Near Surface Environment and H Retention}\label{sec:3}

\begin{figure*}
\includegraphics[width=\linewidth]{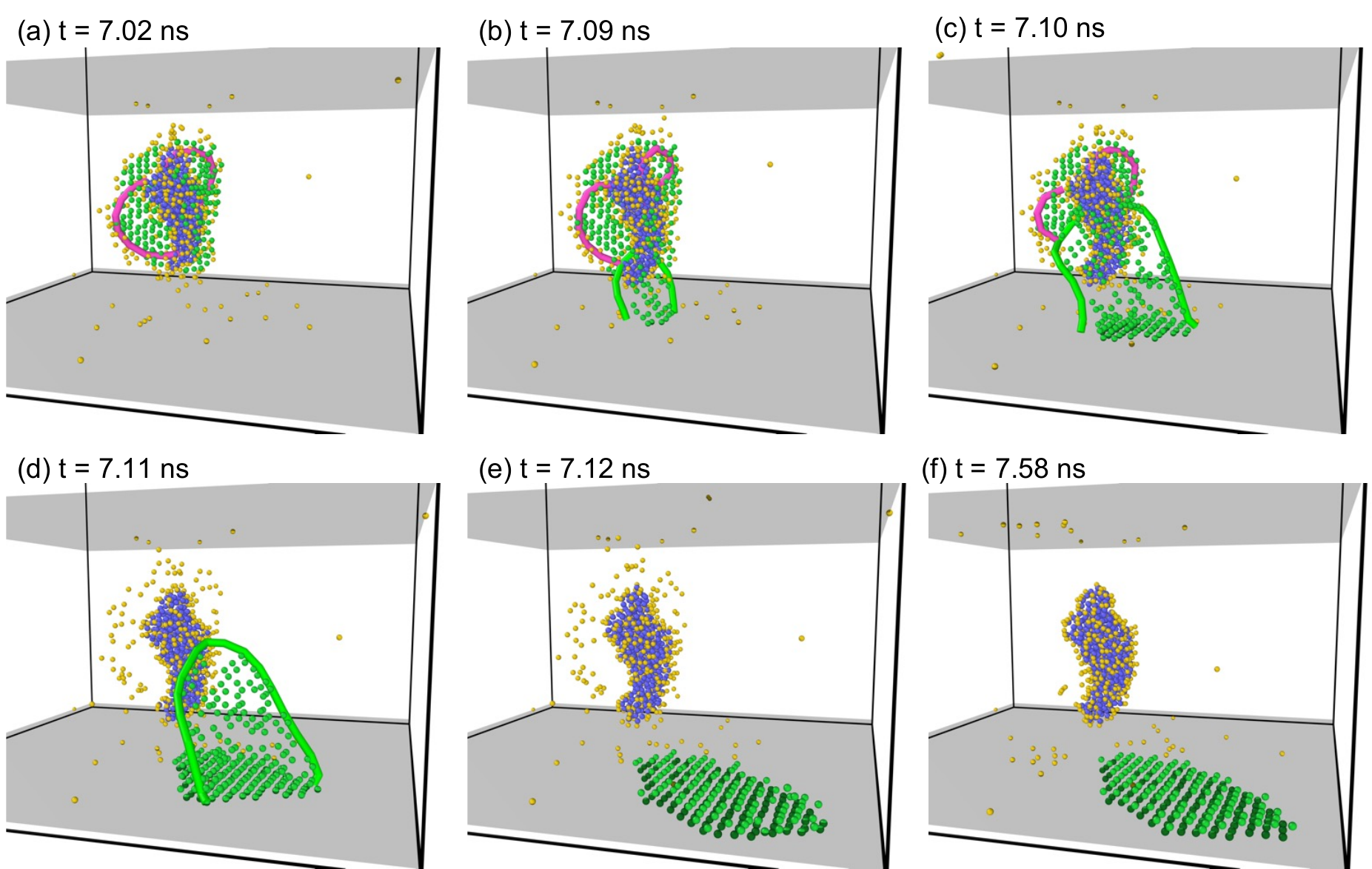}
\caption{\label{fig:Peel} Snapshots of bubble growth simulations showing a He (blue spheres)/ H (yellow spheres) (1:1 ratio) bubbles generated in bulk W with free surfaces (gray planes) having generated W interstitials (green spheres) by Frenkel pair nucleation. Dislocations are also observed of type $<$100$>$ (pink lines) and 1/2$<$111$>$ (green lines).}
\end{figure*}

In a scenario where H does indeed eliminate the formation of 1/2$<$111$>$ dislocation loops, then we would expect the interstitials produced by the bubble to never leave the bubble vicinity since the mechanism to do this is dislocation glide which is only possible with 1/2$<$111$>$ loops. Therefore, it is natural to assume there will be a complex network of dislocation loops and non-loop SIAs (as discussed in section \ref{sec:1}) generated around such a bubble. An example can be seen in figure \ref{fig:Peel}(a) which shows an MD simulation used to generate the data in figure \ref{fig:DislocationVsCount}(1:1). Here we can see a dense ring of W interstitials around the lenticular He:H bubble. This interstitial ring is similar in spirit to that generated when He bubbles are grown at GBs since in that scenario it is thermodynamically favorable to keep the interstitials at the GB close to the bubble \cite{Liu2018,Hatton2024}. Here, though the net effect is similar, the origin is different. In this case the interstitials are \lq stuck' to the bubble partially due to the blocking of SIAs from forming loops but primarily due to the thermodynamic stability of sessile $<$100$>$ loops, both caused by W-H interactions. Thus, the local trapping of these are a consequence of a combined thermodynamic and kinetic effect. The $<100>$ loops are overwhelmingly favored to form, and their sluggish kinetics means there is no mechanism to transport the interstitials away from the bubble.

Visually inspecting the bubble in figure \ref{fig:Peel}(a) we can see that H indeed decorates the periphery of the He bubble as seen in previous studies \cite{Bergstrom2017}. Furthermore, we also find that H decorates the core of the dislocation loops, a finding consistent with DFT studies \cite{DeBacker2017_a,DeBacker2017}; the present study uses different methodologies than both of these previous studies indicating the robustness of this finding. The mechanism of H getting onto dislocation loops is observed to be the transport from the bubble periphery along the dislocation core. This is an effect observed previously in both classical and DFT calculations of dislocation loops, leading to H trapping by dislocation loops, with a dependence on dislocation character \cite{Wang2020,DeBacker2017,Bakaev2017}. This character dependence for H trapping could explain why H is having a varying stabilizing effect on different dislocation loop types.

Furthermore, in figure \ref{fig:Peel}(a) there are a number of H atoms at the free surface. This may imply, quite naturally, that there is a limit on how many H atoms that a He bubble of a given size can accommodate. Presumably after some bubble decoration with H it becomes less energetically favorable for the H to stay at the bubble and therefore some goes into to the bulk and eventually to the free surface. This limit is not known explicitly, however, static calculations conclude that a single He-Vacancy complex can trap 12 H atoms before it is more thermodynamically favorable for the H to be in a interstitial position far from the bubble \cite{Zhou2010,Jiang2010}. Clearly this is not the case in our simulations which implies that this static analysis based on single He-V complexes may not be representative of scenarios concerning larger He bubbles.

The continued evolution of this particular bubble is shown in figure \ref{fig:Peel}(a)-(f) where we see a quite remarkable mechanism. Once the bubble reaches some critical distance from the surface we see (fig.~\ref{fig:Peel}(b)) the nucleation of a 1/2$<$111$>$ dislocation loop (green) which, over the next few ps, \lq peels' \emph{all} of the W interstitials (green spheres) from the bubble's periphery and deposits them at the surface (figs.~\ref{fig:Peel}(c-d)) creating a large plate of new surface atoms (fig.~\ref{fig:Peel}(e)). The dislocation appears to move up each side of the bubble (c-d) collecting the interstitials around the bubble until the loop meets at the top and glides to the free surface.

There are a number of interesting points revealed by this singular event. Notice that before this process took place there were a number of H atoms atoms fixed to the $<$100$>$ loops and W interstitials. However, after the W interstitals are \lq peeled-off' to the surface, the dislocations are not present and the H atoms are released. We see them either move to decorate the bubble or go to the free surface. This mechanism may act as dramatic H release mechanism. Further, the H did not follow the loop to the surface, implying that the H, though bound to the loop, are not able to keep up with the motion of the loop as it peels away. This is consistent with the fact that H is only weakly bound to the 1/2$<$111$>$ loops (or, alternatively, only weakly affect their formation energy). Thus, once the 1/2$<$111$>$ loop forms, the H is no longer bound to the loop and is able to escape the vicinity.

The \lq peeling' mechanism observed here is distinct from the periodic loop-punching mechanisms observed in pure He cases as it requires that the bubble is within some critical distance from the surface in order to stabilize the 1/2$<$111$>$ loops and begin the cascade of W interstitials to the surface. It also represents a possible new mechanism for H retention and release during a combined gas flux. However, more targeted studies will be required to determine if this mechanism is relevant to experimental reality as the simulations here have the caveat that we are directly feeding the bubble with gas - when a network of dislocations loops form, this may slow the arrival of He and/or H \cite{Liu2018, Sandoval2018}. Thus, more work must be done to assess the importance of this mechanism in more realistic scenarios.

\section{Discussion}\label{sec:Disc}

Our atomistic simulations demonstrate that H can induce a change in the energetic favorability of dislocation loops formed from over-pressurized He bubbles. In particular, our work provides strong evidence that mixed He:H bubbles inhibit a bubble's ability to generate glissile 1/2$<$111$>$ loops through a reduction in the formation energy of the competing, but sessile, $<$100$>$ loops. The importance of this finding lies in the effect that He bubbles are generally thought to have on the W surface morphology which is to cause the progressive roughening of the surface by periodic generation and punching of glissile loops \cite{Setyawan2023}. These sessile loops, in contrast, cannot glide to the surface once produced and, in turn, cannot deposit their SIAs to the surface. The result of this is that interstitials stay attached to the bubble and its vicinity and can form a complex network of sessile dislocations as well as non-loop SIAs. The detailed behavior depends on the He:H ratio and is most evident in our simulations at mixtures close to an equal concentration of both gases.

\begin{figure}
\includegraphics[width=\linewidth]{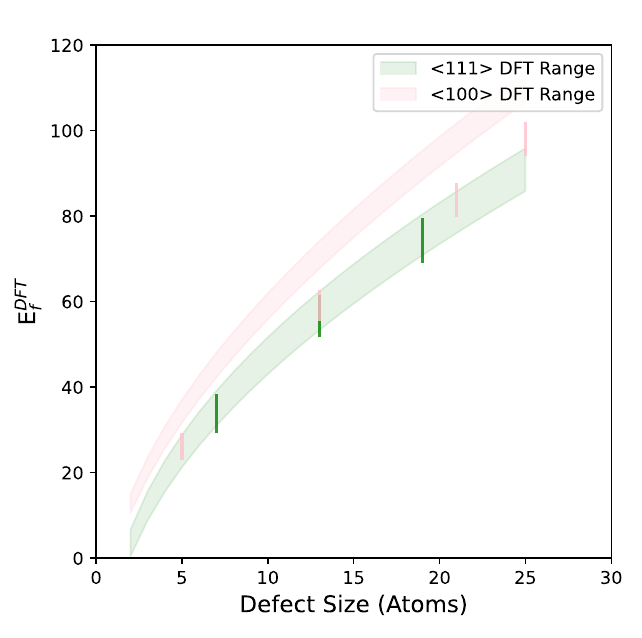}
\caption{\label{fig:FormEn_DFT} Formation energy of isolated dislocation loops with and without H saturation on the loop, as in figure \ref{fig:FormEn_Pot}. The bands highlight ranges of formation energy as predicted from DFT in ref \cite{Ma2020} while the bars are the estimated impact to those formation energies due to the addition of H as revealed from our potential calculations.}
\end{figure}

This conclusion may be tempered by the fact that there is a well-known inconsistency with the interatomic potential used for this work, Marinica's EAM potential \cite{Marinica2013} which is wrapped up into the Gao W-H potential set \cite{Wang2017_b}. The problem is this potential predicts a reversal in the trend in the stability of 1/2$<$111$>$ and $<$100$>$ dislocation loops compared to DFT calculations. That is, with the interatomic potential, $E_f^{<\text{100}>}$ $<$ $E_f^{1/2<\text{111}>}$, as seen in figure \ref{fig:FormEn_Pot} (though only slightly). However, with DFT this trend is flipped such that $E_f^{<\text{100}>}$ $>$ $E_f^{1/2<\text{111}>}$, as can be seen in the shaded regions of figure \ref{fig:FormEn_DFT} which describes the range of energetics described by DFT alongside a Moment Tensor Potential (MTP) developed specifically to capture the DFT dislocation thermodynamics, both taken from \cite{Ma2020}.

To have some sense of how H might impact the stability of loops using DFT, we have shifted values of the computed DFT loop formation energies down by an amount corresponding to the decrease seen in figure \ref{fig:FormEn_Pot} when that dislocation is saturated with H. That is, we consider a scenario in which we use the DFT energetics for the pure loops but combine that with the binding energy of H as determined with the potential. This shift shows that, especially for small dislocation loop sizes, $E_f^{<\text{100}>}$ $\simeq$ $E_f^{1/2<\text{111}>}$. This suggests that the presence of H may indeed impact the energetic favorability of each loop type in DFT and by extension, in experimental reality. Furthermore, there is some experimental evidence to suggest that surface damage is decreased as a result of a mixed flux of He and H to W \cite{Gao2014,Miyamoto2009,Alimov2009,Miyamoto2011} which is consistent with our findings using the interatomic potential. In the end, a primary finding of our MD simulations is that H more significantly impacts the stability of $<$100$>$ loops than it does 1/2$<$111$>$ loops, shifting the balance of the nature of loops punched from bubbles from glissile to sessile. Further DFT calculations may be needed to confirm these conclusions.

A trend that this potential does quantitatively capture is that of H trapping by both the He bubble and the associated dislocation loops. This process is thought to be responsible for the experimentally observed H blistering suppression caused by the blocking of the formation of H$_2$ molecules by the He bubble \cite{Zhou2010}. Further, the presence of the He bubbles in the near surface region is shown to block the permeation of H deeper into W \cite{Markelj2017}. However, in our atomistic study, we do find that if the mixed gas bubble approaches a critical distance from a free surface, the energetic cost of the glissile 1/2$<$111$>$ loop decreases and nucleates, pulling all W interstitials surrounding the bubble to the surface and releasing all H that was previously trapped by the dislocations. This is a new mechanism which must be considered when understanding fuel retention in W. A corollary observation from our simulations is that some H leaves the bubble to decorate the dislocation lines. This is observed mostly at high H concentration indicating that the He bubble is at first a stronger sink for H until some critical point and then the H prefers to decorate the dislocation loops, thus, dislocations \emph{can} represent a pathway to drain bubbles of gas. 

Finally, bubble growth rates chosen in computational studies are known to affect the behavior of the bubbles pressure profile over time such that bubbles grown at higher rate are higher in pressure than at a lower rate \cite{Sandoval2018,Hatton2022}. Future studies could consider the impact of growth rate on the nature of punched dislocations in mixed-gas bubbles. However, we expect that the general conclusion, that H stabilizes $<$100$>$ loops and thus reduces interstitial transport via loop punching to free surfaces, is reliable as it is supported by our thermodynamic calculations.

\section{Conclusion}

Our atomistic study of the role of H on He bubble growth and loop punching behavior indicates that there is a significant impact to the dislocation generation and behavior in mixed He:H gas bubbles compared to pure He bubbles. H does this in multiple compounding ways. Firstly, H in low concentrations affects the ability for W interstitials to form dislocation loops leading to an increase in non-loop SIAs and therefore a reduction in the presence and length of dislocation loops. Secondly, at high H concentrations, H acts to inhibit the mechanisms of W Frenkel pair creation resulting in an overall decrease in SIAs. Finally, and most impactfully, we find a significant decrease in the generation of glissile 1/2$<$111$>$ dislocation loops when a bubble contains an equal mix of He and H. This is found to be due to a crossover in the energetic favorability of dislocation characters caused by H interactions with the dislocation cores. While further work is required to target these effects in experimental studies, we postulate our findings could have significant impact on the predicted damage evolution and fuel retention mechanisms in W in a fusion reactor environment. \\

\begin{acknowledgments}

The authors are grateful for the many fruitful conversations with Brian Wirth during the conception and execution of this research. This research used resources provided by the Los Alamos National Laboratory Institutional Computing Program, which is supported by the U.S. Department of Energy National Nuclear Security Administration under Contract No. 89233218CNA000001.
This project is part of the Scientific Discovery through Advanced Computing (SciDAC) program, and is jointly sponsored by the Fusion Energy Sciences (FES) and Advanced Scientific Computing Research (ASCR) programs within the US Department of Energy, Office of Science. Research supported by the US Department of Energy under DE-AC05-00OR22725. Los Alamos National Laboratory is operated by Triad National Security, LLC, for the National Nuclear Security Administration of U.S. Department of Energy (Contract No. 89233218CNA000001).

\end{acknowledgments}

\bibliography{bibliography}{}

\begin{thebibliography}{69}%
\makeatletter
\providecommand \@ifxundefined [1]{%
 \@ifx{#1\undefined}
}%
\providecommand \@ifnum [1]{%
 \ifnum #1\expandafter \@firstoftwo
 \else \expandafter \@secondoftwo
 \fi
}%
\providecommand \@ifx [1]{%
 \ifx #1\expandafter \@firstoftwo
 \else \expandafter \@secondoftwo
 \fi
}%
\providecommand \natexlab [1]{#1}%
\providecommand \enquote  [1]{``#1''}%
\providecommand \bibnamefont  [1]{#1}%
\providecommand \bibfnamefont [1]{#1}%
\providecommand \citenamefont [1]{#1}%
\providecommand \href@noop [0]{\@secondoftwo}%
\providecommand \href [0]{\begingroup \@sanitize@url \@href}%
\providecommand \@href[1]{\@@startlink{#1}\@@href}%
\providecommand \@@href[1]{\endgroup#1\@@endlink}%
\providecommand \@sanitize@url [0]{\catcode `\\12\catcode `\$12\catcode `\&12\catcode `\#12\catcode `\^12\catcode `\_12\catcode `\%12\relax}%
\providecommand \@@startlink[1]{}%
\providecommand \@@endlink[0]{}%
\providecommand \url  [0]{\begingroup\@sanitize@url \@url }%
\providecommand \@url [1]{\endgroup\@href {#1}{\urlprefix }}%
\providecommand \urlprefix  [0]{URL }%
\providecommand \Eprint [0]{\href }%
\providecommand \doibase [0]{http://dx.doi.org/}%
\providecommand \selectlanguage [0]{\@gobble}%
\providecommand \bibinfo  [0]{\@secondoftwo}%
\providecommand \bibfield  [0]{\@secondoftwo}%
\providecommand \translation [1]{[#1]}%
\providecommand \BibitemOpen [0]{}%
\providecommand \bibitemStop [0]{}%
\providecommand \bibitemNoStop [0]{.\EOS\space}%
\providecommand \EOS [0]{\spacefactor3000\relax}%
\providecommand \BibitemShut  [1]{\csname bibitem#1\endcsname}%
\let\auto@bib@innerbib\@empty
\bibitem [{\citenamefont {Chen}\ \emph {et~al.}(2020)\citenamefont {Chen}, \citenamefont {Wang}, \citenamefont {Li}, \citenamefont {Wang}, \citenamefont {Morgan}, \citenamefont {Xu}, \citenamefont {Chiu},\ and\ \citenamefont {Liu}}]{Chen2020}%
  \BibitemOpen
  \bibfield  {author} {\bibinfo {author} {\bibfnamefont {W.}~\bibnamefont {Chen}}, \bibinfo {author} {\bibfnamefont {X.}~\bibnamefont {Wang}}, \bibinfo {author} {\bibfnamefont {K.}~\bibnamefont {Li}}, \bibinfo {author} {\bibfnamefont {Y.}~\bibnamefont {Wang}}, \bibinfo {author} {\bibfnamefont {T.}~\bibnamefont {Morgan}}, \bibinfo {author} {\bibfnamefont {B.}~\bibnamefont {Xu}}, \bibinfo {author} {\bibfnamefont {Y.}~\bibnamefont {Chiu}}, \ and\ \bibinfo {author} {\bibfnamefont {W.}~\bibnamefont {Liu}},\ }\href@noop {} {\bibfield  {journal} {\bibinfo  {journal} {Scripta Materialia}\ }\textbf {\bibinfo {volume} {187}},\ \bibinfo {pages} {243} (\bibinfo {year} {2020})}\BibitemShut {NoStop}%
\bibitem [{\citenamefont {Iwakiri}\ \emph {et~al.}(2000)\citenamefont {Iwakiri}, \citenamefont {Yasunaga}, \citenamefont {Morishita},\ and\ \citenamefont {Yoshida}}]{Iwakiri2000}%
  \BibitemOpen
  \bibfield  {author} {\bibinfo {author} {\bibfnamefont {H.}~\bibnamefont {Iwakiri}}, \bibinfo {author} {\bibfnamefont {K.}~\bibnamefont {Yasunaga}}, \bibinfo {author} {\bibfnamefont {K.}~\bibnamefont {Morishita}}, \ and\ \bibinfo {author} {\bibfnamefont {N.}~\bibnamefont {Yoshida}},\ }\href@noop {} {\bibfield  {journal} {\bibinfo  {journal} {Journal of nuclear materials}\ }\textbf {\bibinfo {volume} {283}},\ \bibinfo {pages} {1134} (\bibinfo {year} {2000})}\BibitemShut {NoStop}%
\bibitem [{\citenamefont {Sefta}\ \emph {et~al.}(2013)\citenamefont {Sefta}, \citenamefont {Hammond}, \citenamefont {Juslin},\ and\ \citenamefont {Wirth}}]{Sefta2013}%
  \BibitemOpen
  \bibfield  {author} {\bibinfo {author} {\bibfnamefont {F.}~\bibnamefont {Sefta}}, \bibinfo {author} {\bibfnamefont {K.~D.}\ \bibnamefont {Hammond}}, \bibinfo {author} {\bibfnamefont {N.}~\bibnamefont {Juslin}}, \ and\ \bibinfo {author} {\bibfnamefont {B.~D.}\ \bibnamefont {Wirth}},\ }\href@noop {} {\bibfield  {journal} {\bibinfo  {journal} {Nuclear Fusion}\ }\textbf {\bibinfo {volume} {53}},\ \bibinfo {pages} {073015} (\bibinfo {year} {2013})}\BibitemShut {NoStop}%
\bibitem [{\citenamefont {Fikar}\ \emph {et~al.}(2017)\citenamefont {Fikar}, \citenamefont {Gr{\"o}ger},\ and\ \citenamefont {Sch{\"a}ublin}}]{Fikar2017}%
  \BibitemOpen
  \bibfield  {author} {\bibinfo {author} {\bibfnamefont {J.}~\bibnamefont {Fikar}}, \bibinfo {author} {\bibfnamefont {R.}~\bibnamefont {Gr{\"o}ger}}, \ and\ \bibinfo {author} {\bibfnamefont {R.}~\bibnamefont {Sch{\"a}ublin}},\ }\href@noop {} {\bibfield  {journal} {\bibinfo  {journal} {Nuclear Instruments and Methods in Physics Research Section B: Beam Interactions with Materials and Atoms}\ }\textbf {\bibinfo {volume} {393}},\ \bibinfo {pages} {186} (\bibinfo {year} {2017})}\BibitemShut {NoStop}%
\bibitem [{\citenamefont {Sandoval}\ \emph {et~al.}(2018)\citenamefont {Sandoval}, \citenamefont {Perez}, \citenamefont {Uberuaga},\ and\ \citenamefont {Voter}}]{Sandoval2018}%
  \BibitemOpen
  \bibfield  {author} {\bibinfo {author} {\bibfnamefont {L.}~\bibnamefont {Sandoval}}, \bibinfo {author} {\bibfnamefont {D.}~\bibnamefont {Perez}}, \bibinfo {author} {\bibfnamefont {B.~P.}\ \bibnamefont {Uberuaga}}, \ and\ \bibinfo {author} {\bibfnamefont {A.~F.}\ \bibnamefont {Voter}},\ }\href@noop {} {\bibfield  {journal} {\bibinfo  {journal} {Acta Materialia}\ }\textbf {\bibinfo {volume} {159}},\ \bibinfo {pages} {46} (\bibinfo {year} {2018})}\BibitemShut {NoStop}%
\bibitem [{\citenamefont {Hammond}\ \emph {et~al.}(2020)\citenamefont {Hammond}, \citenamefont {Maroudas},\ and\ \citenamefont {Wirth}}]{Hammond2020}%
  \BibitemOpen
  \bibfield  {author} {\bibinfo {author} {\bibfnamefont {K.~D.}\ \bibnamefont {Hammond}}, \bibinfo {author} {\bibfnamefont {D.}~\bibnamefont {Maroudas}}, \ and\ \bibinfo {author} {\bibfnamefont {B.~D.}\ \bibnamefont {Wirth}},\ }\href@noop {} {\bibfield  {journal} {\bibinfo  {journal} {Scientific Reports}\ }\textbf {\bibinfo {volume} {10}},\ \bibinfo {pages} {2192} (\bibinfo {year} {2020})}\BibitemShut {NoStop}%
\bibitem [{\citenamefont {Hammond}\ \emph {et~al.}(2019)\citenamefont {Hammond}, \citenamefont {Naeger}, \citenamefont {Widanagamaachchi}, \citenamefont {Lo}, \citenamefont {Maroudas},\ and\ \citenamefont {Wirth}}]{Hammond2019}%
  \BibitemOpen
  \bibfield  {author} {\bibinfo {author} {\bibfnamefont {K.~D.}\ \bibnamefont {Hammond}}, \bibinfo {author} {\bibfnamefont {I.~V.}\ \bibnamefont {Naeger}}, \bibinfo {author} {\bibfnamefont {W.}~\bibnamefont {Widanagamaachchi}}, \bibinfo {author} {\bibfnamefont {L.-T.}\ \bibnamefont {Lo}}, \bibinfo {author} {\bibfnamefont {D.}~\bibnamefont {Maroudas}}, \ and\ \bibinfo {author} {\bibfnamefont {B.~D.}\ \bibnamefont {Wirth}},\ }\href@noop {} {\bibfield  {journal} {\bibinfo  {journal} {Nuclear Fusion}\ }\textbf {\bibinfo {volume} {59}},\ \bibinfo {pages} {066035} (\bibinfo {year} {2019})}\BibitemShut {NoStop}%
\bibitem [{\citenamefont {Fikar}\ and\ \citenamefont {Gr{\"o}ger}(2015)}]{Fikar2015}%
  \BibitemOpen
  \bibfield  {author} {\bibinfo {author} {\bibfnamefont {J.}~\bibnamefont {Fikar}}\ and\ \bibinfo {author} {\bibfnamefont {R.}~\bibnamefont {Gr{\"o}ger}},\ }\href@noop {} {\bibfield  {journal} {\bibinfo  {journal} {Acta Materialia}\ }\textbf {\bibinfo {volume} {99}},\ \bibinfo {pages} {392} (\bibinfo {year} {2015})}\BibitemShut {NoStop}%
\bibitem [{\citenamefont {Perez}\ \emph {et~al.}(2017)\citenamefont {Perez}, \citenamefont {Sandoval}, \citenamefont {Blondel}, \citenamefont {Wirth}, \citenamefont {Uberuaga},\ and\ \citenamefont {Voter}}]{Perez2017}%
  \BibitemOpen
  \bibfield  {author} {\bibinfo {author} {\bibfnamefont {D.}~\bibnamefont {Perez}}, \bibinfo {author} {\bibfnamefont {L.}~\bibnamefont {Sandoval}}, \bibinfo {author} {\bibfnamefont {S.}~\bibnamefont {Blondel}}, \bibinfo {author} {\bibfnamefont {B.~D.}\ \bibnamefont {Wirth}}, \bibinfo {author} {\bibfnamefont {B.~P.}\ \bibnamefont {Uberuaga}}, \ and\ \bibinfo {author} {\bibfnamefont {A.~F.}\ \bibnamefont {Voter}},\ }\href@noop {} {\bibfield  {journal} {\bibinfo  {journal} {Scientific reports}\ }\textbf {\bibinfo {volume} {7}},\ \bibinfo {pages} {1} (\bibinfo {year} {2017})}\BibitemShut {NoStop}%
\bibitem [{\citenamefont {Sandoval}\ \emph {et~al.}(2015)\citenamefont {Sandoval}, \citenamefont {Perez}, \citenamefont {Uberuaga},\ and\ \citenamefont {Voter}}]{Sandoval2015}%
  \BibitemOpen
  \bibfield  {author} {\bibinfo {author} {\bibfnamefont {L.}~\bibnamefont {Sandoval}}, \bibinfo {author} {\bibfnamefont {D.}~\bibnamefont {Perez}}, \bibinfo {author} {\bibfnamefont {B.~P.}\ \bibnamefont {Uberuaga}}, \ and\ \bibinfo {author} {\bibfnamefont {A.~F.}\ \bibnamefont {Voter}},\ }\href@noop {} {\bibfield  {journal} {\bibinfo  {journal} {Physical review letters}\ }\textbf {\bibinfo {volume} {114}},\ \bibinfo {pages} {105502} (\bibinfo {year} {2015})}\BibitemShut {NoStop}%
\bibitem [{\citenamefont {Ito}\ \emph {et~al.}(2014)\citenamefont {Ito}, \citenamefont {Yoshimoto}, \citenamefont {Saito}, \citenamefont {Takayama},\ and\ \citenamefont {Nakamura}}]{Ito2014}%
  \BibitemOpen
  \bibfield  {author} {\bibinfo {author} {\bibfnamefont {A.~M.}\ \bibnamefont {Ito}}, \bibinfo {author} {\bibfnamefont {Y.}~\bibnamefont {Yoshimoto}}, \bibinfo {author} {\bibfnamefont {S.}~\bibnamefont {Saito}}, \bibinfo {author} {\bibfnamefont {A.}~\bibnamefont {Takayama}}, \ and\ \bibinfo {author} {\bibfnamefont {H.}~\bibnamefont {Nakamura}},\ }\href@noop {} {\bibfield  {journal} {\bibinfo  {journal} {Physica Scripta}\ }\textbf {\bibinfo {volume} {2014}},\ \bibinfo {pages} {014062} (\bibinfo {year} {2014})}\BibitemShut {NoStop}%
\bibitem [{\citenamefont {Dasgupta}\ \emph {et~al.}(2019)\citenamefont {Dasgupta}, \citenamefont {Kolasinski}, \citenamefont {Friddle}, \citenamefont {Du}, \citenamefont {Maroudas},\ and\ \citenamefont {Wirth}}]{Dasgupta2019}%
  \BibitemOpen
  \bibfield  {author} {\bibinfo {author} {\bibfnamefont {D.}~\bibnamefont {Dasgupta}}, \bibinfo {author} {\bibfnamefont {R.~D.}\ \bibnamefont {Kolasinski}}, \bibinfo {author} {\bibfnamefont {R.~W.}\ \bibnamefont {Friddle}}, \bibinfo {author} {\bibfnamefont {L.}~\bibnamefont {Du}}, \bibinfo {author} {\bibfnamefont {D.}~\bibnamefont {Maroudas}}, \ and\ \bibinfo {author} {\bibfnamefont {B.~D.}\ \bibnamefont {Wirth}},\ }\href@noop {} {\bibfield  {journal} {\bibinfo  {journal} {Nuclear Fusion}\ }\textbf {\bibinfo {volume} {59}},\ \bibinfo {pages} {086057} (\bibinfo {year} {2019})}\BibitemShut {NoStop}%
\bibitem [{\citenamefont {Ueda}\ \emph {et~al.}(2013)\citenamefont {Ueda}, \citenamefont {Peng}, \citenamefont {Lee}, \citenamefont {Ohno}, \citenamefont {Kajita}, \citenamefont {Yoshida}, \citenamefont {Doerner}, \citenamefont {De~Temmerman}, \citenamefont {Alimov},\ and\ \citenamefont {Wright}}]{Ueda2013}%
  \BibitemOpen
  \bibfield  {author} {\bibinfo {author} {\bibfnamefont {Y.}~\bibnamefont {Ueda}}, \bibinfo {author} {\bibfnamefont {H.}~\bibnamefont {Peng}}, \bibinfo {author} {\bibfnamefont {H.}~\bibnamefont {Lee}}, \bibinfo {author} {\bibfnamefont {N.}~\bibnamefont {Ohno}}, \bibinfo {author} {\bibfnamefont {S.}~\bibnamefont {Kajita}}, \bibinfo {author} {\bibfnamefont {N.}~\bibnamefont {Yoshida}}, \bibinfo {author} {\bibfnamefont {R.}~\bibnamefont {Doerner}}, \bibinfo {author} {\bibfnamefont {G.}~\bibnamefont {De~Temmerman}}, \bibinfo {author} {\bibfnamefont {V.}~\bibnamefont {Alimov}}, \ and\ \bibinfo {author} {\bibfnamefont {G.}~\bibnamefont {Wright}},\ }\href@noop {} {\bibfield  {journal} {\bibinfo  {journal} {Journal of Nuclear Materials}\ }\textbf {\bibinfo {volume} {442}},\ \bibinfo {pages} {S267} (\bibinfo {year} {2013})}\BibitemShut {NoStop}%
\bibitem [{\citenamefont {Ueda}\ \emph {et~al.}(2014)\citenamefont {Ueda}, \citenamefont {Coenen}, \citenamefont {De~Temmerman}, \citenamefont {Doerner}, \citenamefont {Linke}, \citenamefont {Philipps},\ and\ \citenamefont {Tsitrone}}]{Ueda2014}%
  \BibitemOpen
  \bibfield  {author} {\bibinfo {author} {\bibfnamefont {Y.}~\bibnamefont {Ueda}}, \bibinfo {author} {\bibfnamefont {J.}~\bibnamefont {Coenen}}, \bibinfo {author} {\bibfnamefont {G.}~\bibnamefont {De~Temmerman}}, \bibinfo {author} {\bibfnamefont {R.}~\bibnamefont {Doerner}}, \bibinfo {author} {\bibfnamefont {J.}~\bibnamefont {Linke}}, \bibinfo {author} {\bibfnamefont {V.}~\bibnamefont {Philipps}}, \ and\ \bibinfo {author} {\bibfnamefont {E.}~\bibnamefont {Tsitrone}},\ }\href@noop {} {\bibfield  {journal} {\bibinfo  {journal} {Fusion engineering and design}\ }\textbf {\bibinfo {volume} {89}},\ \bibinfo {pages} {901} (\bibinfo {year} {2014})}\BibitemShut {NoStop}%
\bibitem [{\citenamefont {Wang}\ \emph {et~al.}(2017{\natexlab{a}})\citenamefont {Wang}, \citenamefont {Doerner}, \citenamefont {Baldwin}, \citenamefont {Meyer}, \citenamefont {Bannister}, \citenamefont {Darbal}, \citenamefont {Stroud},\ and\ \citenamefont {Parish}}]{Wang2017_a}%
  \BibitemOpen
  \bibfield  {author} {\bibinfo {author} {\bibfnamefont {K.}~\bibnamefont {Wang}}, \bibinfo {author} {\bibfnamefont {R.}~\bibnamefont {Doerner}}, \bibinfo {author} {\bibfnamefont {M.~J.}\ \bibnamefont {Baldwin}}, \bibinfo {author} {\bibfnamefont {F.~W.}\ \bibnamefont {Meyer}}, \bibinfo {author} {\bibfnamefont {M.~E.}\ \bibnamefont {Bannister}}, \bibinfo {author} {\bibfnamefont {A.}~\bibnamefont {Darbal}}, \bibinfo {author} {\bibfnamefont {R.}~\bibnamefont {Stroud}}, \ and\ \bibinfo {author} {\bibfnamefont {C.~M.}\ \bibnamefont {Parish}},\ }\href@noop {} {\bibfield  {journal} {\bibinfo  {journal} {Scientific reports}\ }\textbf {\bibinfo {volume} {7}},\ \bibinfo {pages} {1} (\bibinfo {year} {2017}{\natexlab{a}})}\BibitemShut {NoStop}%
\bibitem [{\citenamefont {Lasa}\ \emph {et~al.}(2013)\citenamefont {Lasa}, \citenamefont {Henriksson},\ and\ \citenamefont {Nordlund}}]{lasa2013}%
  \BibitemOpen
  \bibfield  {author} {\bibinfo {author} {\bibfnamefont {A.}~\bibnamefont {Lasa}}, \bibinfo {author} {\bibfnamefont {K.}~\bibnamefont {Henriksson}}, \ and\ \bibinfo {author} {\bibfnamefont {K.}~\bibnamefont {Nordlund}},\ }\href@noop {} {\bibfield  {journal} {\bibinfo  {journal} {Nuclear Instruments and Methods in Physics Research Section B: Beam Interactions with Materials and Atoms}\ }\textbf {\bibinfo {volume} {303}},\ \bibinfo {pages} {156} (\bibinfo {year} {2013})}\BibitemShut {NoStop}%
\bibitem [{\citenamefont {Lasa}\ \emph {et~al.}(2014)\citenamefont {Lasa}, \citenamefont {T{\"a}htinen},\ and\ \citenamefont {Nordlund}}]{Lasa2014}%
  \BibitemOpen
  \bibfield  {author} {\bibinfo {author} {\bibfnamefont {A.}~\bibnamefont {Lasa}}, \bibinfo {author} {\bibfnamefont {S.~K.}\ \bibnamefont {T{\"a}htinen}}, \ and\ \bibinfo {author} {\bibfnamefont {K.}~\bibnamefont {Nordlund}},\ }\href@noop {} {\bibfield  {journal} {\bibinfo  {journal} {EPL (Europhysics Letters)}\ }\textbf {\bibinfo {volume} {105}},\ \bibinfo {pages} {25002} (\bibinfo {year} {2014})}\BibitemShut {NoStop}%
\bibitem [{\citenamefont {Chen}\ \emph {et~al.}(2022)\citenamefont {Chen}, \citenamefont {Dasgupta}, \citenamefont {Weerasinghe}, \citenamefont {Hammond}, \citenamefont {Wirth},\ and\ \citenamefont {Maroudas}}]{Chen2022}%
  \BibitemOpen
  \bibfield  {author} {\bibinfo {author} {\bibfnamefont {C.-S.}\ \bibnamefont {Chen}}, \bibinfo {author} {\bibfnamefont {D.}~\bibnamefont {Dasgupta}}, \bibinfo {author} {\bibfnamefont {A.}~\bibnamefont {Weerasinghe}}, \bibinfo {author} {\bibfnamefont {K.~D.}\ \bibnamefont {Hammond}}, \bibinfo {author} {\bibfnamefont {B.~D.}\ \bibnamefont {Wirth}}, \ and\ \bibinfo {author} {\bibfnamefont {D.}~\bibnamefont {Maroudas}},\ }\href@noop {} {\bibfield  {journal} {\bibinfo  {journal} {Nuclear Fusion}\ } (\bibinfo {year} {2022})}\BibitemShut {NoStop}%
\bibitem [{\citenamefont {El-Atwani}\ \emph {et~al.}(2017)\citenamefont {El-Atwani}, \citenamefont {Nathaniel}, \citenamefont {Leff}, \citenamefont {Hattar},\ and\ \citenamefont {Taheri}}]{ElAtwani2017}%
  \BibitemOpen
  \bibfield  {author} {\bibinfo {author} {\bibfnamefont {O.}~\bibnamefont {El-Atwani}}, \bibinfo {author} {\bibfnamefont {J.}~\bibnamefont {Nathaniel}}, \bibinfo {author} {\bibfnamefont {A.~C.}\ \bibnamefont {Leff}}, \bibinfo {author} {\bibfnamefont {K.}~\bibnamefont {Hattar}}, \ and\ \bibinfo {author} {\bibfnamefont {M.}~\bibnamefont {Taheri}},\ }\href@noop {} {\bibfield  {journal} {\bibinfo  {journal} {Scientific reports}\ }\textbf {\bibinfo {volume} {7}},\ \bibinfo {pages} {1} (\bibinfo {year} {2017})}\BibitemShut {NoStop}%
\bibitem [{\citenamefont {El-Atwani}\ \emph {et~al.}(2015)\citenamefont {El-Atwani}, \citenamefont {Suslova}, \citenamefont {Novakowski}, \citenamefont {Hattar}, \citenamefont {Efe}, \citenamefont {Harilal},\ and\ \citenamefont {Hassanein}}]{ElAtwani2015_a}%
  \BibitemOpen
  \bibfield  {author} {\bibinfo {author} {\bibfnamefont {O.}~\bibnamefont {El-Atwani}}, \bibinfo {author} {\bibfnamefont {A.}~\bibnamefont {Suslova}}, \bibinfo {author} {\bibfnamefont {T.}~\bibnamefont {Novakowski}}, \bibinfo {author} {\bibfnamefont {K.}~\bibnamefont {Hattar}}, \bibinfo {author} {\bibfnamefont {M.}~\bibnamefont {Efe}}, \bibinfo {author} {\bibfnamefont {S.}~\bibnamefont {Harilal}}, \ and\ \bibinfo {author} {\bibfnamefont {A.}~\bibnamefont {Hassanein}},\ }\href@noop {} {\bibfield  {journal} {\bibinfo  {journal} {Materials Characterization}\ }\textbf {\bibinfo {volume} {99}},\ \bibinfo {pages} {68} (\bibinfo {year} {2015})}\BibitemShut {NoStop}%
\bibitem [{\citenamefont {Bai}\ \emph {et~al.}(2010)\citenamefont {Bai}, \citenamefont {Voter}, \citenamefont {Hoagland}, \citenamefont {Nastasi},\ and\ \citenamefont {Uberuaga}}]{Bai2010}%
  \BibitemOpen
  \bibfield  {author} {\bibinfo {author} {\bibfnamefont {X.-M.}\ \bibnamefont {Bai}}, \bibinfo {author} {\bibfnamefont {A.~F.}\ \bibnamefont {Voter}}, \bibinfo {author} {\bibfnamefont {R.~G.}\ \bibnamefont {Hoagland}}, \bibinfo {author} {\bibfnamefont {M.}~\bibnamefont {Nastasi}}, \ and\ \bibinfo {author} {\bibfnamefont {B.~P.}\ \bibnamefont {Uberuaga}},\ }\href@noop {} {\bibfield  {journal} {\bibinfo  {journal} {Science}\ }\textbf {\bibinfo {volume} {327}},\ \bibinfo {pages} {1631} (\bibinfo {year} {2010})}\BibitemShut {NoStop}%
\bibitem [{\citenamefont {Uberuaga}\ \emph {et~al.}(2015)\citenamefont {Uberuaga}, \citenamefont {Vernon}, \citenamefont {Martinez},\ and\ \citenamefont {Voter}}]{Uberuaga2015}%
  \BibitemOpen
  \bibfield  {author} {\bibinfo {author} {\bibfnamefont {B.~P.}\ \bibnamefont {Uberuaga}}, \bibinfo {author} {\bibfnamefont {L.~J.}\ \bibnamefont {Vernon}}, \bibinfo {author} {\bibfnamefont {E.}~\bibnamefont {Martinez}}, \ and\ \bibinfo {author} {\bibfnamefont {A.~F.}\ \bibnamefont {Voter}},\ }\href@noop {} {\bibfield  {journal} {\bibinfo  {journal} {Scientific reports}\ }\textbf {\bibinfo {volume} {5}},\ \bibinfo {pages} {1} (\bibinfo {year} {2015})}\BibitemShut {NoStop}%
\bibitem [{\citenamefont {Bai}\ and\ \citenamefont {Uberuaga}(2013)}]{Bai2013}%
  \BibitemOpen
  \bibfield  {author} {\bibinfo {author} {\bibfnamefont {X.-M.}\ \bibnamefont {Bai}}\ and\ \bibinfo {author} {\bibfnamefont {B.~P.}\ \bibnamefont {Uberuaga}},\ }\href@noop {} {\bibfield  {journal} {\bibinfo  {journal} {Jom}\ }\textbf {\bibinfo {volume} {65}},\ \bibinfo {pages} {360} (\bibinfo {year} {2013})}\BibitemShut {NoStop}%
\bibitem [{\citenamefont {Xiao}\ \emph {et~al.}(2016)\citenamefont {Xiao}, \citenamefont {Chu},\ and\ \citenamefont {Duan}}]{Xiao2016}%
  \BibitemOpen
  \bibfield  {author} {\bibinfo {author} {\bibfnamefont {X.}~\bibnamefont {Xiao}}, \bibinfo {author} {\bibfnamefont {H.}~\bibnamefont {Chu}}, \ and\ \bibinfo {author} {\bibfnamefont {H.}~\bibnamefont {Duan}},\ }\href@noop {} {\bibfield  {journal} {\bibinfo  {journal} {Science China Physics, Mechanics \& Astronomy}\ }\textbf {\bibinfo {volume} {59}},\ \bibinfo {pages} {1} (\bibinfo {year} {2016})}\BibitemShut {NoStop}%
\bibitem [{\citenamefont {Hatton}\ \emph {et~al.}(2024)\citenamefont {Hatton}, \citenamefont {Perez}, \citenamefont {Frolov},\ and\ \citenamefont {Uberuaga}}]{Hatton2024}%
  \BibitemOpen
  \bibfield  {author} {\bibinfo {author} {\bibfnamefont {P.}~\bibnamefont {Hatton}}, \bibinfo {author} {\bibfnamefont {D.}~\bibnamefont {Perez}}, \bibinfo {author} {\bibfnamefont {T.}~\bibnamefont {Frolov}}, \ and\ \bibinfo {author} {\bibfnamefont {B.~P.}\ \bibnamefont {Uberuaga}},\ }\href@noop {} {\bibfield  {journal} {\bibinfo  {journal} {Acta Materialia}\ ,\ \bibinfo {pages} {119821}} (\bibinfo {year} {2024})}\BibitemShut {NoStop}%
\bibitem [{\citenamefont {Frolov}\ \emph {et~al.}(2018)\citenamefont {Frolov}, \citenamefont {Zhu}, \citenamefont {Oppelstrup}, \citenamefont {Marian},\ and\ \citenamefont {Rudd}}]{Frolov2018}%
  \BibitemOpen
  \bibfield  {author} {\bibinfo {author} {\bibfnamefont {T.}~\bibnamefont {Frolov}}, \bibinfo {author} {\bibfnamefont {Q.}~\bibnamefont {Zhu}}, \bibinfo {author} {\bibfnamefont {T.}~\bibnamefont {Oppelstrup}}, \bibinfo {author} {\bibfnamefont {J.}~\bibnamefont {Marian}}, \ and\ \bibinfo {author} {\bibfnamefont {R.~E.}\ \bibnamefont {Rudd}},\ }\href {\doibase https://doi.org/10.1016/j.actamat.2018.07.051} {\bibfield  {journal} {\bibinfo  {journal} {Acta Materialia}\ }\textbf {\bibinfo {volume} {159}},\ \bibinfo {pages} {123} (\bibinfo {year} {2018})}\BibitemShut {NoStop}%
\bibitem [{\citenamefont {Lu}\ \emph {et~al.}(2014)\citenamefont {Lu}, \citenamefont {Zhou},\ and\ \citenamefont {Becquart}}]{Lu2014}%
  \BibitemOpen
  \bibfield  {author} {\bibinfo {author} {\bibfnamefont {G.-H.}\ \bibnamefont {Lu}}, \bibinfo {author} {\bibfnamefont {H.-B.}\ \bibnamefont {Zhou}}, \ and\ \bibinfo {author} {\bibfnamefont {C.~S.}\ \bibnamefont {Becquart}},\ }\href@noop {} {\bibfield  {journal} {\bibinfo  {journal} {Nuclear Fusion}\ }\textbf {\bibinfo {volume} {54}},\ \bibinfo {pages} {086001} (\bibinfo {year} {2014})}\BibitemShut {NoStop}%
\bibitem [{\citenamefont {Liu}\ \emph {et~al.}(2009{\natexlab{a}})\citenamefont {Liu}, \citenamefont {Zhang}, \citenamefont {Luo},\ and\ \citenamefont {Lu}}]{Liu2009}%
  \BibitemOpen
  \bibfield  {author} {\bibinfo {author} {\bibfnamefont {Y.-L.}\ \bibnamefont {Liu}}, \bibinfo {author} {\bibfnamefont {Y.}~\bibnamefont {Zhang}}, \bibinfo {author} {\bibfnamefont {G.-N.}\ \bibnamefont {Luo}}, \ and\ \bibinfo {author} {\bibfnamefont {G.-H.}\ \bibnamefont {Lu}},\ }\href@noop {} {\bibfield  {journal} {\bibinfo  {journal} {Journal of Nuclear Materials}\ }\textbf {\bibinfo {volume} {390}},\ \bibinfo {pages} {1032} (\bibinfo {year} {2009}{\natexlab{a}})}\BibitemShut {NoStop}%
\bibitem [{\citenamefont {Johnson}\ and\ \citenamefont {Carter}(2010)}]{Johnson2010}%
  \BibitemOpen
  \bibfield  {author} {\bibinfo {author} {\bibfnamefont {D.~F.}\ \bibnamefont {Johnson}}\ and\ \bibinfo {author} {\bibfnamefont {E.~A.}\ \bibnamefont {Carter}},\ }\href@noop {} {\bibfield  {journal} {\bibinfo  {journal} {Journal of Materials Research}\ }\textbf {\bibinfo {volume} {25}},\ \bibinfo {pages} {315} (\bibinfo {year} {2010})}\BibitemShut {NoStop}%
\bibitem [{\citenamefont {Heinola}\ and\ \citenamefont {Ahlgren}(2010)}]{Heinola2010}%
  \BibitemOpen
  \bibfield  {author} {\bibinfo {author} {\bibfnamefont {K.}~\bibnamefont {Heinola}}\ and\ \bibinfo {author} {\bibfnamefont {T.}~\bibnamefont {Ahlgren}},\ }\href@noop {} {\bibfield  {journal} {\bibinfo  {journal} {Journal of Applied physics}\ }\textbf {\bibinfo {volume} {107}} (\bibinfo {year} {2010})}\BibitemShut {NoStop}%
\bibitem [{\citenamefont {Frauenfelder}(1969)}]{Frauenfelder1969}%
  \BibitemOpen
  \bibfield  {author} {\bibinfo {author} {\bibfnamefont {R.}~\bibnamefont {Frauenfelder}},\ }\href@noop {} {\bibfield  {journal} {\bibinfo  {journal} {Journal of Vacuum Science and Technology}\ }\textbf {\bibinfo {volume} {6}},\ \bibinfo {pages} {388} (\bibinfo {year} {1969})}\BibitemShut {NoStop}%
\bibitem [{\citenamefont {Duan}\ \emph {et~al.}(2010)\citenamefont {Duan}, \citenamefont {Liu}, \citenamefont {Zhou}, \citenamefont {Zhang}, \citenamefont {Jin}, \citenamefont {Lu},\ and\ \citenamefont {Luo}}]{Duan2010}%
  \BibitemOpen
  \bibfield  {author} {\bibinfo {author} {\bibfnamefont {C.}~\bibnamefont {Duan}}, \bibinfo {author} {\bibfnamefont {Y.-L.}\ \bibnamefont {Liu}}, \bibinfo {author} {\bibfnamefont {H.-B.}\ \bibnamefont {Zhou}}, \bibinfo {author} {\bibfnamefont {Y.}~\bibnamefont {Zhang}}, \bibinfo {author} {\bibfnamefont {S.}~\bibnamefont {Jin}}, \bibinfo {author} {\bibfnamefont {G.-H.}\ \bibnamefont {Lu}}, \ and\ \bibinfo {author} {\bibfnamefont {G.-N.}\ \bibnamefont {Luo}},\ }\href@noop {} {\bibfield  {journal} {\bibinfo  {journal} {Journal of nuclear materials}\ }\textbf {\bibinfo {volume} {404}},\ \bibinfo {pages} {109} (\bibinfo {year} {2010})}\BibitemShut {NoStop}%
\bibitem [{\citenamefont {Liu}\ \emph {et~al.}(2009{\natexlab{b}})\citenamefont {Liu}, \citenamefont {Zhang}, \citenamefont {Zhou}, \citenamefont {Lu}, \citenamefont {Liu},\ and\ \citenamefont {Luo}}]{Liu2009_a}%
  \BibitemOpen
  \bibfield  {author} {\bibinfo {author} {\bibfnamefont {Y.-L.}\ \bibnamefont {Liu}}, \bibinfo {author} {\bibfnamefont {Y.}~\bibnamefont {Zhang}}, \bibinfo {author} {\bibfnamefont {H.-B.}\ \bibnamefont {Zhou}}, \bibinfo {author} {\bibfnamefont {G.-H.}\ \bibnamefont {Lu}}, \bibinfo {author} {\bibfnamefont {F.}~\bibnamefont {Liu}}, \ and\ \bibinfo {author} {\bibfnamefont {G.-N.}\ \bibnamefont {Luo}},\ }\href@noop {} {\bibfield  {journal} {\bibinfo  {journal} {Physical Review B}\ }\textbf {\bibinfo {volume} {79}},\ \bibinfo {pages} {172103} (\bibinfo {year} {2009}{\natexlab{b}})}\BibitemShut {NoStop}%
\bibitem [{\citenamefont {Ohsawa}\ \emph {et~al.}(2010)\citenamefont {Ohsawa}, \citenamefont {Goto}, \citenamefont {Yamakami}, \citenamefont {Yamaguchi},\ and\ \citenamefont {Yagi}}]{Ohsawa2010}%
  \BibitemOpen
  \bibfield  {author} {\bibinfo {author} {\bibfnamefont {K.}~\bibnamefont {Ohsawa}}, \bibinfo {author} {\bibfnamefont {J.}~\bibnamefont {Goto}}, \bibinfo {author} {\bibfnamefont {M.}~\bibnamefont {Yamakami}}, \bibinfo {author} {\bibfnamefont {M.}~\bibnamefont {Yamaguchi}}, \ and\ \bibinfo {author} {\bibfnamefont {M.}~\bibnamefont {Yagi}},\ }\href@noop {} {\bibfield  {journal} {\bibinfo  {journal} {Physical Review B}\ }\textbf {\bibinfo {volume} {82}},\ \bibinfo {pages} {184117} (\bibinfo {year} {2010})}\BibitemShut {NoStop}%
\bibitem [{\citenamefont {Ohsawa}\ \emph {et~al.}(2012)\citenamefont {Ohsawa}, \citenamefont {Eguchi}, \citenamefont {Watanabe}, \citenamefont {Yamaguchi},\ and\ \citenamefont {Yagi}}]{Ohsawa2012}%
  \BibitemOpen
  \bibfield  {author} {\bibinfo {author} {\bibfnamefont {K.}~\bibnamefont {Ohsawa}}, \bibinfo {author} {\bibfnamefont {K.}~\bibnamefont {Eguchi}}, \bibinfo {author} {\bibfnamefont {H.}~\bibnamefont {Watanabe}}, \bibinfo {author} {\bibfnamefont {M.}~\bibnamefont {Yamaguchi}}, \ and\ \bibinfo {author} {\bibfnamefont {M.}~\bibnamefont {Yagi}},\ }\href@noop {} {\bibfield  {journal} {\bibinfo  {journal} {Physical Review B}\ }\textbf {\bibinfo {volume} {85}},\ \bibinfo {pages} {094102} (\bibinfo {year} {2012})}\BibitemShut {NoStop}%
\bibitem [{\citenamefont {Ye}\ \emph {et~al.}(2003)\citenamefont {Ye}, \citenamefont {Kanehara}, \citenamefont {Fukuta}, \citenamefont {Ohno},\ and\ \citenamefont {Takamura}}]{Ye2003}%
  \BibitemOpen
  \bibfield  {author} {\bibinfo {author} {\bibfnamefont {M.}~\bibnamefont {Ye}}, \bibinfo {author} {\bibfnamefont {H.}~\bibnamefont {Kanehara}}, \bibinfo {author} {\bibfnamefont {S.}~\bibnamefont {Fukuta}}, \bibinfo {author} {\bibfnamefont {N.}~\bibnamefont {Ohno}}, \ and\ \bibinfo {author} {\bibfnamefont {S.}~\bibnamefont {Takamura}},\ }\href@noop {} {\bibfield  {journal} {\bibinfo  {journal} {Journal of nuclear materials}\ }\textbf {\bibinfo {volume} {313}},\ \bibinfo {pages} {72} (\bibinfo {year} {2003})}\BibitemShut {NoStop}%
\bibitem [{\citenamefont {Tanabe}(2014)}]{Tanabe2014}%
  \BibitemOpen
  \bibfield  {author} {\bibinfo {author} {\bibfnamefont {T.}~\bibnamefont {Tanabe}},\ }\href@noop {} {\bibfield  {journal} {\bibinfo  {journal} {Physica Scripta}\ }\textbf {\bibinfo {volume} {2014}},\ \bibinfo {pages} {014044} (\bibinfo {year} {2014})}\BibitemShut {NoStop}%
\bibitem [{\citenamefont {Hatton}\ \emph {et~al.}(2019)\citenamefont {Hatton}, \citenamefont {Goddard}, \citenamefont {Smith}, \citenamefont {Abbas}, \citenamefont {Potamialis}, \citenamefont {Greenhalgh},\ and\ \citenamefont {Walls}}]{Hatton2019}%
  \BibitemOpen
  \bibfield  {author} {\bibinfo {author} {\bibfnamefont {P.}~\bibnamefont {Hatton}}, \bibinfo {author} {\bibfnamefont {P.}~\bibnamefont {Goddard}}, \bibinfo {author} {\bibfnamefont {R.}~\bibnamefont {Smith}}, \bibinfo {author} {\bibfnamefont {A.}~\bibnamefont {Abbas}}, \bibinfo {author} {\bibfnamefont {C.}~\bibnamefont {Potamialis}}, \bibinfo {author} {\bibfnamefont {R.}~\bibnamefont {Greenhalgh}}, \ and\ \bibinfo {author} {\bibfnamefont {J.}~\bibnamefont {Walls}},\ }\href@noop {} {\bibfield  {journal} {\bibinfo  {journal} {Thin Solid Films}\ }\textbf {\bibinfo {volume} {692}},\ \bibinfo {pages} {137614} (\bibinfo {year} {2019})}\BibitemShut {NoStop}%
\bibitem [{\citenamefont {Hatton}\ \emph {et~al.}(2020)\citenamefont {Hatton}, \citenamefont {Abbas}, \citenamefont {Kaminski}, \citenamefont {Yilmaz}, \citenamefont {Watts}, \citenamefont {Walls}, \citenamefont {Goddard},\ and\ \citenamefont {Smith}}]{Hatton2020}%
  \BibitemOpen
  \bibfield  {author} {\bibinfo {author} {\bibfnamefont {P.}~\bibnamefont {Hatton}}, \bibinfo {author} {\bibfnamefont {A.}~\bibnamefont {Abbas}}, \bibinfo {author} {\bibfnamefont {P.}~\bibnamefont {Kaminski}}, \bibinfo {author} {\bibfnamefont {S.}~\bibnamefont {Yilmaz}}, \bibinfo {author} {\bibfnamefont {M.}~\bibnamefont {Watts}}, \bibinfo {author} {\bibfnamefont {M.}~\bibnamefont {Walls}}, \bibinfo {author} {\bibfnamefont {P.}~\bibnamefont {Goddard}}, \ and\ \bibinfo {author} {\bibfnamefont {R.}~\bibnamefont {Smith}},\ }\href@noop {} {\bibfield  {journal} {\bibinfo  {journal} {Proceedings of the Royal Society A}\ }\textbf {\bibinfo {volume} {476}},\ \bibinfo {pages} {20200056} (\bibinfo {year} {2020})}\BibitemShut {NoStop}%
\bibitem [{\citenamefont {Yu}\ \emph {et~al.}(2014)\citenamefont {Yu}, \citenamefont {Shu}, \citenamefont {Liu},\ and\ \citenamefont {Lu}}]{Yu2014}%
  \BibitemOpen
  \bibfield  {author} {\bibinfo {author} {\bibfnamefont {Y.}~\bibnamefont {Yu}}, \bibinfo {author} {\bibfnamefont {X.}~\bibnamefont {Shu}}, \bibinfo {author} {\bibfnamefont {Y.-N.}\ \bibnamefont {Liu}}, \ and\ \bibinfo {author} {\bibfnamefont {G.-H.}\ \bibnamefont {Lu}},\ }\href@noop {} {\bibfield  {journal} {\bibinfo  {journal} {Journal of Nuclear Materials}\ }\textbf {\bibinfo {volume} {455}},\ \bibinfo {pages} {91} (\bibinfo {year} {2014})}\BibitemShut {NoStop}%
\bibitem [{\citenamefont {Wang}\ \emph {et~al.}(2017{\natexlab{b}})\citenamefont {Wang}, \citenamefont {Niu},\ and\ \citenamefont {Wang}}]{Wang2017}%
  \BibitemOpen
  \bibfield  {author} {\bibinfo {author} {\bibfnamefont {X.-X.}\ \bibnamefont {Wang}}, \bibinfo {author} {\bibfnamefont {L.-L.}\ \bibnamefont {Niu}}, \ and\ \bibinfo {author} {\bibfnamefont {S.}~\bibnamefont {Wang}},\ }\href@noop {} {\bibfield  {journal} {\bibinfo  {journal} {Journal of Nuclear Materials}\ }\textbf {\bibinfo {volume} {487}},\ \bibinfo {pages} {158} (\bibinfo {year} {2017}{\natexlab{b}})}\BibitemShut {NoStop}%
\bibitem [{\citenamefont {Mathew}\ \emph {et~al.}(2022)\citenamefont {Mathew}, \citenamefont {Perez}, \citenamefont {Suk}, \citenamefont {Uberuaga},\ and\ \citenamefont {Martinez}}]{Mathew2022}%
  \BibitemOpen
  \bibfield  {author} {\bibinfo {author} {\bibfnamefont {N.}~\bibnamefont {Mathew}}, \bibinfo {author} {\bibfnamefont {D.}~\bibnamefont {Perez}}, \bibinfo {author} {\bibfnamefont {W.}~\bibnamefont {Suk}}, \bibinfo {author} {\bibfnamefont {B.~P.}\ \bibnamefont {Uberuaga}}, \ and\ \bibinfo {author} {\bibfnamefont {E.}~\bibnamefont {Martinez}},\ }\href@noop {} {\bibfield  {journal} {\bibinfo  {journal} {Nuclear Fusion}\ } (\bibinfo {year} {2022})}\BibitemShut {NoStop}%
\bibitem [{\citenamefont {Shi}\ \emph {et~al.}(2023)\citenamefont {Shi}, \citenamefont {Li}, \citenamefont {Li}, \citenamefont {Liu}, \citenamefont {Fan}, \citenamefont {Peng}, \citenamefont {Liang}, \citenamefont {Jin},\ and\ \citenamefont {Lu}}]{Shi2023}%
  \BibitemOpen
  \bibfield  {author} {\bibinfo {author} {\bibfnamefont {J.}~\bibnamefont {Shi}}, \bibinfo {author} {\bibfnamefont {B.}~\bibnamefont {Li}}, \bibinfo {author} {\bibfnamefont {L.}~\bibnamefont {Li}}, \bibinfo {author} {\bibfnamefont {Y.}~\bibnamefont {Liu}}, \bibinfo {author} {\bibfnamefont {X.}~\bibnamefont {Fan}}, \bibinfo {author} {\bibfnamefont {Q.}~\bibnamefont {Peng}}, \bibinfo {author} {\bibfnamefont {L.}~\bibnamefont {Liang}}, \bibinfo {author} {\bibfnamefont {S.}~\bibnamefont {Jin}}, \ and\ \bibinfo {author} {\bibfnamefont {G.-H.}\ \bibnamefont {Lu}},\ }\href@noop {} {\bibfield  {journal} {\bibinfo  {journal} {Fusion Engineering and Design}\ }\textbf {\bibinfo {volume} {197}},\ \bibinfo {pages} {114030} (\bibinfo {year} {2023})}\BibitemShut {NoStop}%
\bibitem [{\citenamefont {Ding}\ \emph {et~al.}(2022)\citenamefont {Ding}, \citenamefont {Yu}, \citenamefont {Lin}, \citenamefont {Zhao}, \citenamefont {Xiao}, \citenamefont {Vinogradov}, \citenamefont {Qiao}, \citenamefont {Ortiz}, \citenamefont {He},\ and\ \citenamefont {Zhang}}]{Ding2022}%
  \BibitemOpen
  \bibfield  {author} {\bibinfo {author} {\bibfnamefont {Y.}~\bibnamefont {Ding}}, \bibinfo {author} {\bibfnamefont {H.}~\bibnamefont {Yu}}, \bibinfo {author} {\bibfnamefont {M.}~\bibnamefont {Lin}}, \bibinfo {author} {\bibfnamefont {K.}~\bibnamefont {Zhao}}, \bibinfo {author} {\bibfnamefont {S.}~\bibnamefont {Xiao}}, \bibinfo {author} {\bibfnamefont {A.}~\bibnamefont {Vinogradov}}, \bibinfo {author} {\bibfnamefont {L.}~\bibnamefont {Qiao}}, \bibinfo {author} {\bibfnamefont {M.}~\bibnamefont {Ortiz}}, \bibinfo {author} {\bibfnamefont {J.}~\bibnamefont {He}}, \ and\ \bibinfo {author} {\bibfnamefont {Z.}~\bibnamefont {Zhang}},\ }\href@noop {} {\bibfield  {journal} {\bibinfo  {journal} {Acta Materialia}\ }\textbf {\bibinfo {volume} {239}},\ \bibinfo {pages} {118279} (\bibinfo {year} {2022})}\BibitemShut {NoStop}%
\bibitem [{\citenamefont {Nishijima}\ \emph {et~al.}(2005)\citenamefont {Nishijima}, \citenamefont {Sugimoto}, \citenamefont {Iwakiri}, \citenamefont {Ye}, \citenamefont {Ohno}, \citenamefont {Yoshida},\ and\ \citenamefont {Takamura}}]{Nishijima2005}%
  \BibitemOpen
  \bibfield  {author} {\bibinfo {author} {\bibfnamefont {D.}~\bibnamefont {Nishijima}}, \bibinfo {author} {\bibfnamefont {T.}~\bibnamefont {Sugimoto}}, \bibinfo {author} {\bibfnamefont {H.}~\bibnamefont {Iwakiri}}, \bibinfo {author} {\bibfnamefont {M.}~\bibnamefont {Ye}}, \bibinfo {author} {\bibfnamefont {N.}~\bibnamefont {Ohno}}, \bibinfo {author} {\bibfnamefont {N.}~\bibnamefont {Yoshida}}, \ and\ \bibinfo {author} {\bibfnamefont {S.}~\bibnamefont {Takamura}},\ }\href@noop {} {\bibfield  {journal} {\bibinfo  {journal} {Journal of nuclear materials}\ }\textbf {\bibinfo {volume} {337}},\ \bibinfo {pages} {927} (\bibinfo {year} {2005})}\BibitemShut {NoStop}%
\bibitem [{\citenamefont {Bergstrom}\ \emph {et~al.}(2017)\citenamefont {Bergstrom}, \citenamefont {Cusentino},\ and\ \citenamefont {Wirth}}]{Bergstrom2017}%
  \BibitemOpen
  \bibfield  {author} {\bibinfo {author} {\bibfnamefont {Z.}~\bibnamefont {Bergstrom}}, \bibinfo {author} {\bibfnamefont {M.}~\bibnamefont {Cusentino}}, \ and\ \bibinfo {author} {\bibfnamefont {B.}~\bibnamefont {Wirth}},\ }\href@noop {} {\bibfield  {journal} {\bibinfo  {journal} {Fusion Science and Technology}\ }\textbf {\bibinfo {volume} {71}},\ \bibinfo {pages} {122} (\bibinfo {year} {2017})}\BibitemShut {NoStop}%
\bibitem [{\citenamefont {Gao}\ \emph {et~al.}(2014)\citenamefont {Gao}, \citenamefont {Von~Toussaint}, \citenamefont {Jacob}, \citenamefont {Balden},\ and\ \citenamefont {Manhard}}]{Gao2014}%
  \BibitemOpen
  \bibfield  {author} {\bibinfo {author} {\bibfnamefont {L.}~\bibnamefont {Gao}}, \bibinfo {author} {\bibfnamefont {U.}~\bibnamefont {Von~Toussaint}}, \bibinfo {author} {\bibfnamefont {W.}~\bibnamefont {Jacob}}, \bibinfo {author} {\bibfnamefont {M.}~\bibnamefont {Balden}}, \ and\ \bibinfo {author} {\bibfnamefont {A.}~\bibnamefont {Manhard}},\ }\href@noop {} {\bibfield  {journal} {\bibinfo  {journal} {Nuclear Fusion}\ }\textbf {\bibinfo {volume} {54}},\ \bibinfo {pages} {122003} (\bibinfo {year} {2014})}\BibitemShut {NoStop}%
\bibitem [{\citenamefont {Miyamoto}\ \emph {et~al.}(2009)\citenamefont {Miyamoto}, \citenamefont {Nishijima}, \citenamefont {Ueda}, \citenamefont {Doerner}, \citenamefont {Kurishita}, \citenamefont {Baldwin}, \citenamefont {Morito}, \citenamefont {Ono},\ and\ \citenamefont {Hanna}}]{Miyamoto2009}%
  \BibitemOpen
  \bibfield  {author} {\bibinfo {author} {\bibfnamefont {M.}~\bibnamefont {Miyamoto}}, \bibinfo {author} {\bibfnamefont {D.}~\bibnamefont {Nishijima}}, \bibinfo {author} {\bibfnamefont {Y.}~\bibnamefont {Ueda}}, \bibinfo {author} {\bibfnamefont {R.}~\bibnamefont {Doerner}}, \bibinfo {author} {\bibfnamefont {H.}~\bibnamefont {Kurishita}}, \bibinfo {author} {\bibfnamefont {M.}~\bibnamefont {Baldwin}}, \bibinfo {author} {\bibfnamefont {S.}~\bibnamefont {Morito}}, \bibinfo {author} {\bibfnamefont {K.}~\bibnamefont {Ono}}, \ and\ \bibinfo {author} {\bibfnamefont {J.}~\bibnamefont {Hanna}},\ }\href@noop {} {\bibfield  {journal} {\bibinfo  {journal} {Nuclear Fusion}\ }\textbf {\bibinfo {volume} {49}},\ \bibinfo {pages} {065035} (\bibinfo {year} {2009})}\BibitemShut {NoStop}%
\bibitem [{\citenamefont {Alimov}\ \emph {et~al.}(2009)\citenamefont {Alimov}, \citenamefont {Shu}, \citenamefont {Roth}, \citenamefont {Sugiyama}, \citenamefont {Lindig}, \citenamefont {Balden}, \citenamefont {Isobe},\ and\ \citenamefont {Yamanishi}}]{Alimov2009}%
  \BibitemOpen
  \bibfield  {author} {\bibinfo {author} {\bibfnamefont {V.~K.}\ \bibnamefont {Alimov}}, \bibinfo {author} {\bibfnamefont {W.}~\bibnamefont {Shu}}, \bibinfo {author} {\bibfnamefont {J.}~\bibnamefont {Roth}}, \bibinfo {author} {\bibfnamefont {K.}~\bibnamefont {Sugiyama}}, \bibinfo {author} {\bibfnamefont {S.}~\bibnamefont {Lindig}}, \bibinfo {author} {\bibfnamefont {M.}~\bibnamefont {Balden}}, \bibinfo {author} {\bibfnamefont {K.}~\bibnamefont {Isobe}}, \ and\ \bibinfo {author} {\bibfnamefont {T.}~\bibnamefont {Yamanishi}},\ }\href@noop {} {\bibfield  {journal} {\bibinfo  {journal} {Physica Scripta}\ }\textbf {\bibinfo {volume} {2009}},\ \bibinfo {pages} {014048} (\bibinfo {year} {2009})}\BibitemShut {NoStop}%
\bibitem [{\citenamefont {Miyamoto}\ \emph {et~al.}(2011)\citenamefont {Miyamoto}, \citenamefont {Nishijima}, \citenamefont {Baldwin}, \citenamefont {Doerner}, \citenamefont {Ueda}, \citenamefont {Yasunaga}, \citenamefont {Yoshida},\ and\ \citenamefont {Ono}}]{Miyamoto2011}%
  \BibitemOpen
  \bibfield  {author} {\bibinfo {author} {\bibfnamefont {M.}~\bibnamefont {Miyamoto}}, \bibinfo {author} {\bibfnamefont {D.}~\bibnamefont {Nishijima}}, \bibinfo {author} {\bibfnamefont {M.}~\bibnamefont {Baldwin}}, \bibinfo {author} {\bibfnamefont {R.}~\bibnamefont {Doerner}}, \bibinfo {author} {\bibfnamefont {Y.}~\bibnamefont {Ueda}}, \bibinfo {author} {\bibfnamefont {K.}~\bibnamefont {Yasunaga}}, \bibinfo {author} {\bibfnamefont {N.}~\bibnamefont {Yoshida}}, \ and\ \bibinfo {author} {\bibfnamefont {K.}~\bibnamefont {Ono}},\ }\href@noop {} {\bibfield  {journal} {\bibinfo  {journal} {Journal of Nuclear Materials}\ }\textbf {\bibinfo {volume} {415}},\ \bibinfo {pages} {S657} (\bibinfo {year} {2011})}\BibitemShut {NoStop}%
\bibitem [{\citenamefont {Thompson}\ \emph {et~al.}(2022)\citenamefont {Thompson}, \citenamefont {Aktulga}, \citenamefont {Berger}, \citenamefont {Bolintineanu}, \citenamefont {Brown}, \citenamefont {Crozier}, \citenamefont {{in 't Veld}}, \citenamefont {Kohlmeyer}, \citenamefont {Moore}, \citenamefont {Nguyen}, \citenamefont {Shan}, \citenamefont {Stevens}, \citenamefont {Tranchida}, \citenamefont {Trott},\ and\ \citenamefont {Plimpton}}]{LAMMPS2022}%
  \BibitemOpen
  \bibfield  {author} {\bibinfo {author} {\bibfnamefont {A.~P.}\ \bibnamefont {Thompson}}, \bibinfo {author} {\bibfnamefont {H.~M.}\ \bibnamefont {Aktulga}}, \bibinfo {author} {\bibfnamefont {R.}~\bibnamefont {Berger}}, \bibinfo {author} {\bibfnamefont {D.~S.}\ \bibnamefont {Bolintineanu}}, \bibinfo {author} {\bibfnamefont {W.~M.}\ \bibnamefont {Brown}}, \bibinfo {author} {\bibfnamefont {P.~S.}\ \bibnamefont {Crozier}}, \bibinfo {author} {\bibfnamefont {P.~J.}\ \bibnamefont {{in 't Veld}}}, \bibinfo {author} {\bibfnamefont {A.}~\bibnamefont {Kohlmeyer}}, \bibinfo {author} {\bibfnamefont {S.~G.}\ \bibnamefont {Moore}}, \bibinfo {author} {\bibfnamefont {T.~D.}\ \bibnamefont {Nguyen}}, \bibinfo {author} {\bibfnamefont {R.}~\bibnamefont {Shan}}, \bibinfo {author} {\bibfnamefont {M.~J.}\ \bibnamefont {Stevens}}, \bibinfo {author} {\bibfnamefont {J.}~\bibnamefont {Tranchida}}, \bibinfo {author} {\bibfnamefont {C.}~\bibnamefont {Trott}}, \ and\ \bibinfo {author} {\bibfnamefont {S.~J.}\ \bibnamefont {Plimpton}},\ }\href {\doibase https://doi.org/10.1016/j.cpc.2021.108171} {\bibfield  {journal} {\bibinfo  {journal} {Computer Physics Communications}\ }\textbf {\bibinfo {volume} {271}},\ \bibinfo {pages} {108171} (\bibinfo {year} {2022})}\BibitemShut {NoStop}%
\bibitem [{\citenamefont {Wang}\ \emph {et~al.}(2017{\natexlab{c}})\citenamefont {Wang}, \citenamefont {Shu}, \citenamefont {Lu},\ and\ \citenamefont {Gao}}]{Wang2017_b}%
  \BibitemOpen
  \bibfield  {author} {\bibinfo {author} {\bibfnamefont {L.-F.}\ \bibnamefont {Wang}}, \bibinfo {author} {\bibfnamefont {X.}~\bibnamefont {Shu}}, \bibinfo {author} {\bibfnamefont {G.-H.}\ \bibnamefont {Lu}}, \ and\ \bibinfo {author} {\bibfnamefont {F.}~\bibnamefont {Gao}},\ }\href@noop {} {\bibfield  {journal} {\bibinfo  {journal} {Journal of Physics: Condensed Matter}\ }\textbf {\bibinfo {volume} {29}},\ \bibinfo {pages} {435401} (\bibinfo {year} {2017}{\natexlab{c}})}\BibitemShut {NoStop}%
\bibitem [{\citenamefont {Marinica}\ \emph {et~al.}(2013)\citenamefont {Marinica}, \citenamefont {Ventelon}, \citenamefont {Gilbert}, \citenamefont {Proville}, \citenamefont {Dudarev}, \citenamefont {Marian}, \citenamefont {Bencteux},\ and\ \citenamefont {Willaime}}]{Marinica2013}%
  \BibitemOpen
  \bibfield  {author} {\bibinfo {author} {\bibfnamefont {M.-C.}\ \bibnamefont {Marinica}}, \bibinfo {author} {\bibfnamefont {L.}~\bibnamefont {Ventelon}}, \bibinfo {author} {\bibfnamefont {M.}~\bibnamefont {Gilbert}}, \bibinfo {author} {\bibfnamefont {L.}~\bibnamefont {Proville}}, \bibinfo {author} {\bibfnamefont {S.}~\bibnamefont {Dudarev}}, \bibinfo {author} {\bibfnamefont {J.}~\bibnamefont {Marian}}, \bibinfo {author} {\bibfnamefont {G.}~\bibnamefont {Bencteux}}, \ and\ \bibinfo {author} {\bibfnamefont {F.}~\bibnamefont {Willaime}},\ }\href@noop {} {\bibfield  {journal} {\bibinfo  {journal} {Journal of Physics: Condensed Matter}\ }\textbf {\bibinfo {volume} {25}},\ \bibinfo {pages} {395502} (\bibinfo {year} {2013})}\BibitemShut {NoStop}%
\bibitem [{\citenamefont {Juslin}\ and\ \citenamefont {Wirth}(2013)}]{Juslin2013}%
  \BibitemOpen
  \bibfield  {author} {\bibinfo {author} {\bibfnamefont {N.}~\bibnamefont {Juslin}}\ and\ \bibinfo {author} {\bibfnamefont {B.}~\bibnamefont {Wirth}},\ }\href@noop {} {\bibfield  {journal} {\bibinfo  {journal} {Journal of Nuclear Materials}\ }\textbf {\bibinfo {volume} {432}},\ \bibinfo {pages} {61} (\bibinfo {year} {2013})}\BibitemShut {NoStop}%
\bibitem [{\citenamefont {Stukowski}(2010)}]{Stukowski2010}%
  \BibitemOpen
  \bibfield  {author} {\bibinfo {author} {\bibfnamefont {A.}~\bibnamefont {Stukowski}},\ }\href {\doibase {10.1088/0965-0393/18/1/015012}} {\bibfield  {journal} {\bibinfo  {journal} {Modelling and Simulation in Materials Science and Engineering}\ }\textbf {\bibinfo {volume} {{18}}} (\bibinfo {year} {{2010}}),\ {10.1088/0965-0393/18/1/015012}}\BibitemShut {NoStop}%
\bibitem [{\citenamefont {Yi}\ \emph {et~al.}(2015)\citenamefont {Yi}, \citenamefont {Jenkins}, \citenamefont {Hattar}, \citenamefont {Edmondson},\ and\ \citenamefont {Roberts}}]{Yi2015}%
  \BibitemOpen
  \bibfield  {author} {\bibinfo {author} {\bibfnamefont {X.}~\bibnamefont {Yi}}, \bibinfo {author} {\bibfnamefont {M.~L.}\ \bibnamefont {Jenkins}}, \bibinfo {author} {\bibfnamefont {K.}~\bibnamefont {Hattar}}, \bibinfo {author} {\bibfnamefont {P.~D.}\ \bibnamefont {Edmondson}}, \ and\ \bibinfo {author} {\bibfnamefont {S.~G.}\ \bibnamefont {Roberts}},\ }\href@noop {} {\bibfield  {journal} {\bibinfo  {journal} {Acta Materialia}\ }\textbf {\bibinfo {volume} {92}},\ \bibinfo {pages} {163} (\bibinfo {year} {2015})}\BibitemShut {NoStop}%
\bibitem [{\citenamefont {Yi}\ \emph {et~al.}(2016)\citenamefont {Yi}, \citenamefont {Jenkins}, \citenamefont {Kirk}, \citenamefont {Zhou},\ and\ \citenamefont {Roberts}}]{Yi2016}%
  \BibitemOpen
  \bibfield  {author} {\bibinfo {author} {\bibfnamefont {X.}~\bibnamefont {Yi}}, \bibinfo {author} {\bibfnamefont {M.~L.}\ \bibnamefont {Jenkins}}, \bibinfo {author} {\bibfnamefont {M.~A.}\ \bibnamefont {Kirk}}, \bibinfo {author} {\bibfnamefont {Z.}~\bibnamefont {Zhou}}, \ and\ \bibinfo {author} {\bibfnamefont {S.~G.}\ \bibnamefont {Roberts}},\ }\href@noop {} {\bibfield  {journal} {\bibinfo  {journal} {Acta Materialia}\ }\textbf {\bibinfo {volume} {112}},\ \bibinfo {pages} {105} (\bibinfo {year} {2016})}\BibitemShut {NoStop}%
\bibitem [{\citenamefont {Hull}\ and\ \citenamefont {Bacon}(2001)}]{Hull2001}%
  \BibitemOpen
  \bibfield  {author} {\bibinfo {author} {\bibfnamefont {D.}~\bibnamefont {Hull}}\ and\ \bibinfo {author} {\bibfnamefont {D.~J.}\ \bibnamefont {Bacon}},\ }\href@noop {} {\emph {\bibinfo {title} {Introduction to dislocations}}}\ (\bibinfo  {publisher} {Butterworth-Heinemann},\ \bibinfo {year} {2001})\BibitemShut {NoStop}%
\bibitem [{\citenamefont {Setyawan}\ \emph {et~al.}(2023)\citenamefont {Setyawan}, \citenamefont {Dasgupta}, \citenamefont {Blondel}, \citenamefont {Nandipati}, \citenamefont {Hammond}, \citenamefont {Maroudas},\ and\ \citenamefont {Wirth}}]{Setyawan2023}%
  \BibitemOpen
  \bibfield  {author} {\bibinfo {author} {\bibfnamefont {W.}~\bibnamefont {Setyawan}}, \bibinfo {author} {\bibfnamefont {D.}~\bibnamefont {Dasgupta}}, \bibinfo {author} {\bibfnamefont {S.}~\bibnamefont {Blondel}}, \bibinfo {author} {\bibfnamefont {G.}~\bibnamefont {Nandipati}}, \bibinfo {author} {\bibfnamefont {K.~D.}\ \bibnamefont {Hammond}}, \bibinfo {author} {\bibfnamefont {D.}~\bibnamefont {Maroudas}}, \ and\ \bibinfo {author} {\bibfnamefont {B.~D.}\ \bibnamefont {Wirth}},\ }\href@noop {} {\bibfield  {journal} {\bibinfo  {journal} {Scientific Reports}\ }\textbf {\bibinfo {volume} {13}},\ \bibinfo {pages} {9601} (\bibinfo {year} {2023})}\BibitemShut {NoStop}%
\bibitem [{\citenamefont {De~Backer}\ \emph {et~al.}(2017{\natexlab{a}})\citenamefont {De~Backer}, \citenamefont {Mason}, \citenamefont {Domain}, \citenamefont {Nguyen-Manh}, \citenamefont {Marinica}, \citenamefont {Ventelon}, \citenamefont {Becquart},\ and\ \citenamefont {Dudarev}}]{DeBacker2017}%
  \BibitemOpen
  \bibfield  {author} {\bibinfo {author} {\bibfnamefont {A.}~\bibnamefont {De~Backer}}, \bibinfo {author} {\bibfnamefont {D.~R.}\ \bibnamefont {Mason}}, \bibinfo {author} {\bibfnamefont {C.}~\bibnamefont {Domain}}, \bibinfo {author} {\bibfnamefont {D.}~\bibnamefont {Nguyen-Manh}}, \bibinfo {author} {\bibfnamefont {M.-C.}\ \bibnamefont {Marinica}}, \bibinfo {author} {\bibfnamefont {L.}~\bibnamefont {Ventelon}}, \bibinfo {author} {\bibfnamefont {C.}~\bibnamefont {Becquart}}, \ and\ \bibinfo {author} {\bibfnamefont {S.~L.}\ \bibnamefont {Dudarev}},\ }\href@noop {} {\bibfield  {journal} {\bibinfo  {journal} {Nuclear Fusion}\ }\textbf {\bibinfo {volume} {58}},\ \bibinfo {pages} {016006} (\bibinfo {year} {2017}{\natexlab{a}})}\BibitemShut {NoStop}%
\bibitem [{\citenamefont {Liu}\ \emph {et~al.}(2018)\citenamefont {Liu}, \citenamefont {Uberuaga}, \citenamefont {Perez},\ and\ \citenamefont {Voter}}]{Liu2018}%
  \BibitemOpen
  \bibfield  {author} {\bibinfo {author} {\bibfnamefont {X.-Y.}\ \bibnamefont {Liu}}, \bibinfo {author} {\bibfnamefont {B.~P.}\ \bibnamefont {Uberuaga}}, \bibinfo {author} {\bibfnamefont {D.}~\bibnamefont {Perez}}, \ and\ \bibinfo {author} {\bibfnamefont {A.~F.}\ \bibnamefont {Voter}},\ }\href {\doibase 10.1080/21663831.2018.1494637} {\bibfield  {journal} {\bibinfo  {journal} {Materials Research Letters}\ }\textbf {\bibinfo {volume} {6}},\ \bibinfo {pages} {522} (\bibinfo {year} {2018})}\BibitemShut {NoStop}%
\bibitem [{\citenamefont {De~Backer}\ \emph {et~al.}(2017{\natexlab{b}})\citenamefont {De~Backer}, \citenamefont {Mason}, \citenamefont {Domain}, \citenamefont {Nguyen-Manh}, \citenamefont {Marinica}, \citenamefont {Ventelon}, \citenamefont {Becquart},\ and\ \citenamefont {Dudarev}}]{DeBacker2017_a}%
  \BibitemOpen
  \bibfield  {author} {\bibinfo {author} {\bibfnamefont {A.}~\bibnamefont {De~Backer}}, \bibinfo {author} {\bibfnamefont {D.~R.}\ \bibnamefont {Mason}}, \bibinfo {author} {\bibfnamefont {C.}~\bibnamefont {Domain}}, \bibinfo {author} {\bibfnamefont {D.}~\bibnamefont {Nguyen-Manh}}, \bibinfo {author} {\bibfnamefont {M.-C.}\ \bibnamefont {Marinica}}, \bibinfo {author} {\bibfnamefont {L.}~\bibnamefont {Ventelon}}, \bibinfo {author} {\bibfnamefont {C.~S.}\ \bibnamefont {Becquart}}, \ and\ \bibinfo {author} {\bibfnamefont {S.~L.}\ \bibnamefont {Dudarev}},\ }\href@noop {} {\bibfield  {journal} {\bibinfo  {journal} {Physica Scripta}\ }\textbf {\bibinfo {volume} {2017}},\ \bibinfo {pages} {014073} (\bibinfo {year} {2017}{\natexlab{b}})}\BibitemShut {NoStop}%
\bibitem [{\citenamefont {Wang}\ \emph {et~al.}(2020)\citenamefont {Wang}, \citenamefont {Shu}, \citenamefont {Lin}, \citenamefont {Lu},\ and\ \citenamefont {Song}}]{Wang2020}%
  \BibitemOpen
  \bibfield  {author} {\bibinfo {author} {\bibfnamefont {L.-F.}\ \bibnamefont {Wang}}, \bibinfo {author} {\bibfnamefont {X.}~\bibnamefont {Shu}}, \bibinfo {author} {\bibfnamefont {D.-Y.}\ \bibnamefont {Lin}}, \bibinfo {author} {\bibfnamefont {G.-H.}\ \bibnamefont {Lu}}, \ and\ \bibinfo {author} {\bibfnamefont {H.-F.}\ \bibnamefont {Song}},\ }\href@noop {} {\bibfield  {journal} {\bibinfo  {journal} {International Journal of Hydrogen Energy}\ }\textbf {\bibinfo {volume} {45}},\ \bibinfo {pages} {822} (\bibinfo {year} {2020})}\BibitemShut {NoStop}%
\bibitem [{\citenamefont {Bakaev}\ \emph {et~al.}(2017)\citenamefont {Bakaev}, \citenamefont {Grigorev}, \citenamefont {Terentyev}, \citenamefont {Bakaeva}, \citenamefont {Zhurkin},\ and\ \citenamefont {Mastrikov}}]{Bakaev2017}%
  \BibitemOpen
  \bibfield  {author} {\bibinfo {author} {\bibfnamefont {A.}~\bibnamefont {Bakaev}}, \bibinfo {author} {\bibfnamefont {P.}~\bibnamefont {Grigorev}}, \bibinfo {author} {\bibfnamefont {D.}~\bibnamefont {Terentyev}}, \bibinfo {author} {\bibfnamefont {A.}~\bibnamefont {Bakaeva}}, \bibinfo {author} {\bibfnamefont {E.}~\bibnamefont {Zhurkin}}, \ and\ \bibinfo {author} {\bibfnamefont {Y.~A.}\ \bibnamefont {Mastrikov}},\ }\href@noop {} {\bibfield  {journal} {\bibinfo  {journal} {Nuclear Fusion}\ }\textbf {\bibinfo {volume} {57}},\ \bibinfo {pages} {126040} (\bibinfo {year} {2017})}\BibitemShut {NoStop}%
\bibitem [{\citenamefont {Zhou}\ \emph {et~al.}(2010)\citenamefont {Zhou}, \citenamefont {Liu}, \citenamefont {Jin}, \citenamefont {Zhang}, \citenamefont {Luo},\ and\ \citenamefont {Lu}}]{Zhou2010}%
  \BibitemOpen
  \bibfield  {author} {\bibinfo {author} {\bibfnamefont {H.-B.}\ \bibnamefont {Zhou}}, \bibinfo {author} {\bibfnamefont {Y.-L.}\ \bibnamefont {Liu}}, \bibinfo {author} {\bibfnamefont {S.}~\bibnamefont {Jin}}, \bibinfo {author} {\bibfnamefont {Y.}~\bibnamefont {Zhang}}, \bibinfo {author} {\bibfnamefont {G.-N.}\ \bibnamefont {Luo}}, \ and\ \bibinfo {author} {\bibfnamefont {G.-H.}\ \bibnamefont {Lu}},\ }\href@noop {} {\bibfield  {journal} {\bibinfo  {journal} {Nuclear Fusion}\ }\textbf {\bibinfo {volume} {50}},\ \bibinfo {pages} {115010} (\bibinfo {year} {2010})}\BibitemShut {NoStop}%
\bibitem [{\citenamefont {Jiang}\ \emph {et~al.}(2010)\citenamefont {Jiang}, \citenamefont {Wan},\ and\ \citenamefont {Geng}}]{Jiang2010}%
  \BibitemOpen
  \bibfield  {author} {\bibinfo {author} {\bibfnamefont {B.}~\bibnamefont {Jiang}}, \bibinfo {author} {\bibfnamefont {F.}~\bibnamefont {Wan}}, \ and\ \bibinfo {author} {\bibfnamefont {W.}~\bibnamefont {Geng}},\ }\href@noop {} {\bibfield  {journal} {\bibinfo  {journal} {Physical Review B}\ }\textbf {\bibinfo {volume} {81}},\ \bibinfo {pages} {134112} (\bibinfo {year} {2010})}\BibitemShut {NoStop}%
\bibitem [{\citenamefont {Ma}\ \emph {et~al.}(2020)\citenamefont {Ma}, \citenamefont {Mason},\ and\ \citenamefont {Dudarev}}]{Ma2020}%
  \BibitemOpen
  \bibfield  {author} {\bibinfo {author} {\bibfnamefont {P.-W.}\ \bibnamefont {Ma}}, \bibinfo {author} {\bibfnamefont {D.}~\bibnamefont {Mason}}, \ and\ \bibinfo {author} {\bibfnamefont {S.}~\bibnamefont {Dudarev}},\ }\href@noop {} {\bibfield  {journal} {\bibinfo  {journal} {Physical Review Materials}\ }\textbf {\bibinfo {volume} {4}},\ \bibinfo {pages} {103609} (\bibinfo {year} {2020})}\BibitemShut {NoStop}%
\bibitem [{\citenamefont {Markelj}\ \emph {et~al.}(2017)\citenamefont {Markelj}, \citenamefont {Schwarz-Selinger},\ and\ \citenamefont {Zalo{\v{z}}nik}}]{Markelj2017}%
  \BibitemOpen
  \bibfield  {author} {\bibinfo {author} {\bibfnamefont {S.}~\bibnamefont {Markelj}}, \bibinfo {author} {\bibfnamefont {T.}~\bibnamefont {Schwarz-Selinger}}, \ and\ \bibinfo {author} {\bibfnamefont {A.}~\bibnamefont {Zalo{\v{z}}nik}},\ }\href@noop {} {\bibfield  {journal} {\bibinfo  {journal} {Nuclear Fusion}\ }\textbf {\bibinfo {volume} {57}},\ \bibinfo {pages} {064002} (\bibinfo {year} {2017})}\BibitemShut {NoStop}%
\bibitem [{\citenamefont {Hatton}\ \emph {et~al.}(2022)\citenamefont {Hatton}, \citenamefont {Hatton}, \citenamefont {Perez},\ and\ \citenamefont {Uberuaga}}]{Hatton2022}%
  \BibitemOpen
  \bibfield  {author} {\bibinfo {author} {\bibfnamefont {P.}~\bibnamefont {Hatton}}, \bibinfo {author} {\bibfnamefont {M.}~\bibnamefont {Hatton}}, \bibinfo {author} {\bibfnamefont {D.}~\bibnamefont {Perez}}, \ and\ \bibinfo {author} {\bibfnamefont {B.~P.}\ \bibnamefont {Uberuaga}},\ }\href@noop {} {\bibfield  {journal} {\bibinfo  {journal} {MRS Communications}\ }\textbf {\bibinfo {volume} {12}},\ \bibinfo {pages} {1103} (\bibinfo {year} {2022})}\BibitemShut {NoStop}%
\end{thebibliography}%

\end{document}